\documentclass[a4paper,12pt]{article}
\usepackage[left=1.1in,right=1.1in, top=1.1in]{geometry}
\usepackage{graphicx}
\usepackage{amsmath}
\usepackage{caption}
\usepackage{url}
\usepackage{subcaption}
\usepackage{bm}
\usepackage[english]{babel}
\usepackage{array}
\newcolumntype{L}[1]{>{\raggedright\let\newline\\\arraybackslash\hspace{0pt}}m{#1}}
\newcolumntype{C}[1]{>{\centering\let\newline\\\arraybackslash\hspace{0pt}}m{#1}}
\newcolumntype{R}[1]{>{\raggedleft\let\newline\\\arraybackslash\hspace{0pt}}m{#1}}
\begin{document}

\title{\sc 
	Global Estimation of Neonatal Mortality using a Bayesian Hierarchical Splines Regression Model\footnote{The authors would like to thank  Danzhen You, Lucia Hug, Simon Ejdemyr, Jon Pedersen, Jing Liu and Jin Rou New for their work on constructing the database. We are also grateful all members of the (Technical Advisory Group of the) United Nations Inter-agency Group for Child Mortality Estimation for helpful feedback and discussions on the model. 
}
}
\author{ Monica Alexander\footnote{\texttt{monicaalexander@berkeley.edu}} \\  \emph{ University of California, Berkeley} \and Leontine Alkema\footnote{\texttt{leontinealkema@gmail.com}}\\  \emph{ University of Massachusetts, Amherst } 
}
\date{}

\maketitle

\begin{abstract}
In recent years, much of the focus in monitoring child mortality has been on assessing changes in the under-five mortality rate (U5MR). However, as the U5MR decreases, the share of neonatal deaths (within the first month) tends to increase, warranting increased efforts in monitoring this indicator in addition to the U5MR.  A Bayesian splines regression model is presented for estimating neonatal mortality rates (NMR) for all countries. In the model, the relationship between NMR and U5MR is assessed and used to inform estimates, and spline regression models are used to capture country-specific trends. As such, the resulting NMR estimates incorporate trends in overall child mortality while also capturing data-driven trends. The model is fitted to 195 countries using the database from the United Nations Interagency Group for Child Mortality Estimation, producing estimates from 1990, or earlier if data are available, until 2015. The results suggest that, above a U5MR of 34 deaths per 1000 live births, at the global level, a 1 per cent increase in the U5MR leads to a 0.6 per cent decrease in the ratio of NMR to U5MR. Below a U5MR of 34 deaths per 1000 live births, the proportion of deaths under-five that are neonatal is constant at around 54 per cent. However, the relationship between U5MR and NMR varies across countries. The model has now been adopted by the United Nations Inter-agency Group for Child Mortality Estimation.

\end{abstract}

\setlength\parindent{0pt} 
\setlength{\parskip}{\baselineskip}%
\clearpage

\section{Introduction}
In order to evaluate a country's progress in reducing child mortality, it is important to obtain accurate estimates; be able to project mortality levels; and have some indication of the uncertainty in the estimates and projections. In practice, obtaining reliable mortality estimates is often most difficult in developing countries where mortality is relatively high, well-functioning vital registration systems are lacking and the data that are available are often subject to large sampling errors and/or of poor quality. This situation calls for the use of statistical models to help estimate underlying mortality trends.

In recent years, much of the focus on monitoring child mortality has been on assessing changes in the under-five mortality rate (U5MR), which refers to the number of deaths before the age of five per 1000 live births. The focus was driven by Millennium Development Goal (MDG) 4, which called for a two-thirds reduction in under-five mortality between 1990 and 2015. A report on MDG progress released in 2015 by the United Nations showed that, although this target was not met in most regions of the world, notable progress has been made (UN 2015). The global U5MR is less than half of its level in 1990, and despite population growth in developing regions, the number of deaths of children under five has declined. 

As the U5MR decreases, the share of neonatal deaths, i.e.\ deaths occurring in the first month, tend to increase. Globally, the estimated share of under-five deaths that were neonatal in 2015 was 45 per cent, a 13 per cent increase from 1990 (IGME 2015). Indeed, in most regions of the world, the majority of under-five deaths are neonatal; for example, the share is 56 per cent in Developed regions; 51 per cent in Latin America and the Caribbean; and 54 per cent in Western Asia. The share is still less than 50 per cent, however, where the U5MR is relatively high: in Sub-Saharan Africa, the share is only 34 per cent. 

The neonatal equivalent to the U5MR is the neonatal mortality rate (NMR), which is defined as the number of neonatal deaths per 1,000 live births. The increasing importance of neonatal deaths in over child mortality has warranted increased efforts in monitoring NMR in addition to the U5MR (e.g. Lawn et al.\ 2004; \ Mekonnen et al.\ 2013; Lozano et al.\ 2011; Bhutta et al.\ 2010). The United Nations Inter-agency Group for Child Mortality Estimation (IGME) publishes estimates of NMR for all 195 UN member countries (IGME 2015). IGME uses a statistical model to obtain estimates for countries without high quality vital registration data, with U5MR as a predictor (Oestergaard~et~al.~2011). While the method has worked well to capture the main trends in the NMR, it has some disadvantages. Most notably, trends in NMR within a country are driven by the U5MR trends, rather than being driven specifically by the NMR data. 

In this paper, we present a new model for estimating NMR for countries worldwide, overcoming some of the concerns with the current IGME model using a Bayesian hierarchical model framework. From the point of view of modeling mortality levels across countries, a Bayesian approach offers an intuitive way to share information across different countries and time points, and a data model can incorporate different sources of error into the estimates. Increases in computation speed as well as the development of numerical methods have enabled a more widespread use of the Bayesian approach in many fields, including population estimation and forecasting (e.g.\ Girosi and King 2007; Raftery et al.\ 2012; Alkema and New 2014; Schmertmann et al.\ 2014). In this application, the proposed Bayesian model is flexible enough to be used to estimate the NMR in any country, regardless of the amount and sources of data available. Results were produced for 195 countries for at least the years 1990--2015, which covers the MDG period of interest, using a dataset with almost 5,000 observations from various data sources. 

The remainder of the paper is structured as follows. Firstly, the dataset and model are summarized in the next two sections. Some key results are then highlighted, including model validation results, followed by a discussion of the work and possible future avenues. Additional details about the model are provided in the Appendix. 

\section{Data}
Data on NMR are derived from either vital registration (VR) systems; sample vital registration (SVR) systems; or survey data. Data for a particular country may come from one or several of these sources, and the source may vary over time. Table \ref{tbledata} summarizes the availability of data by source type. 

Data from VR systems are derived directly from the registered births and deaths in a country. The observed NMR for a particular country and year is the number of registered deaths within the first month divided by the number of live births. SVR systems refer to vital registration statistics that are collected on a representative sample of the broader population. 

As well as using civil registration systems, NMR observations can also be derived from data collected in surveys. During a survey, a mother is asked to list a full history of all births (and possible deaths) of her children. A retrospective series of NMR observations can then be derived using the birth histories (Pedersen~and~Liu~2012). For the majority of survey data series (72 per cent), microdata are available and it is possible to estimate the sampling error associated with each of the observations. For the remaining 28 per cent,  there is not enough information to calculate the sampling errors from the data and values are imputed (see Methods section). All rates, ratios of rates, and corresponding standard errors were calculated from the survey microdata using the software `CMRJack' (Pedersen~and~Liu~2012). Optimized time series are calculated such that all estimates have a coefficient of variation of less than 10 per cent. 

The majority of survey data come from Demographic and Health Surveys (DHS) (Table~\ref{tbledata}). The category `Other DHS' refers to non-standard DHS, that is, Special Interim and National DHS, Malaria Indicator Surveys, AIDS Indicator Surveys and World Fertility Surveys (WFS). National DHS are surveys in DHS format that are run by a national agency, rather than the external DHS agency.  The Multiple Indicator Cluster Survey (MICS), developed by UNICEF in 1990, was originally designed to address trends in goals from the World Summit for Children, and has since focused on assessing progress towards the relevant MDG indicators. The `Other' category includes surveys such as the Pan-Arab Project for Family Health and the Reproductive Health Surveys. 

Data availability varies by country and by year. For most developed countries, a full time series of VR data exists. For other countries with VR data, the time series is often incomplete and is supported by other sources of data. Of the 105 countries where VR data are available, 44 countries have incomplete VR times series. For some smaller countries with VR data, observations were recombined to avoid issues with erratic trends due to large stochastic variance. See Appendix \ref{VRrecalc} for more details. 
SVR data are only available for Bangladesh, China and South Africa. Most developing countries have no vital registration systems and so observations of the NMR are derived entirely from surveys. A total of twelve countries had no available data.  

In terms of data inclusion, we follow the same inclusion exclusion rules as the UN IGME-estimated U5MR (IGME 2015). These exclusion rules are based on external information which suggests some NMR observations are unreliable, due to, for example, poor survey quality or under coverage of VR systems. A total of 16 per cent of the 4,678 observations were excluded. 

\begin{table}[h!]
\caption{Summary of the NMR data availability by source type and whether or not sampling errors were reported. The totals include observations that were excluded from the estimation.}
\label{tbledata}
\begin{tabular}{l|R{1.5cm}|R{2cm}|R{1.5cm}|R{2cm}}
\hline
Source                              &{No.\ of Series} & {No.\ of Countries}  &{No.\ of Obs} & {No. of Country-years}\\ \hline
VR                                  & 105                                &105& 2607                                    &2607\\
SVR                                 & 3                                  &3& 79                                      &78\\
DHS (with sampling errors)          & 239                    &81            & 1212                                   &934 \\
DHS (without sampling errors)       & 16                      & 15          & 50                                     &48 \\
Other DHS (with sampling errors)    & 52                      &       42    & 251                                     &251\\
Other DHS (without sampling errors) & 26                      & 21         & 78                                      &75\\
MICS (with sampling errors)         & 16                              & 14 & 81                                      &73\\
MICS (without sampling errors)      & 12                              &  12  & 49                                      &46\\
Others (with sampling errors)       & 24                                 &16& 119                                     &111\\
Others (without sampling errors)    & 72                                 &36& 152                                     &151\\ \hline
\end{tabular}
\end{table}

\begin{figure} \centering 

\begin{subfigure}[b]{0.47\textwidth}  \includegraphics[width=\textwidth, page=130]{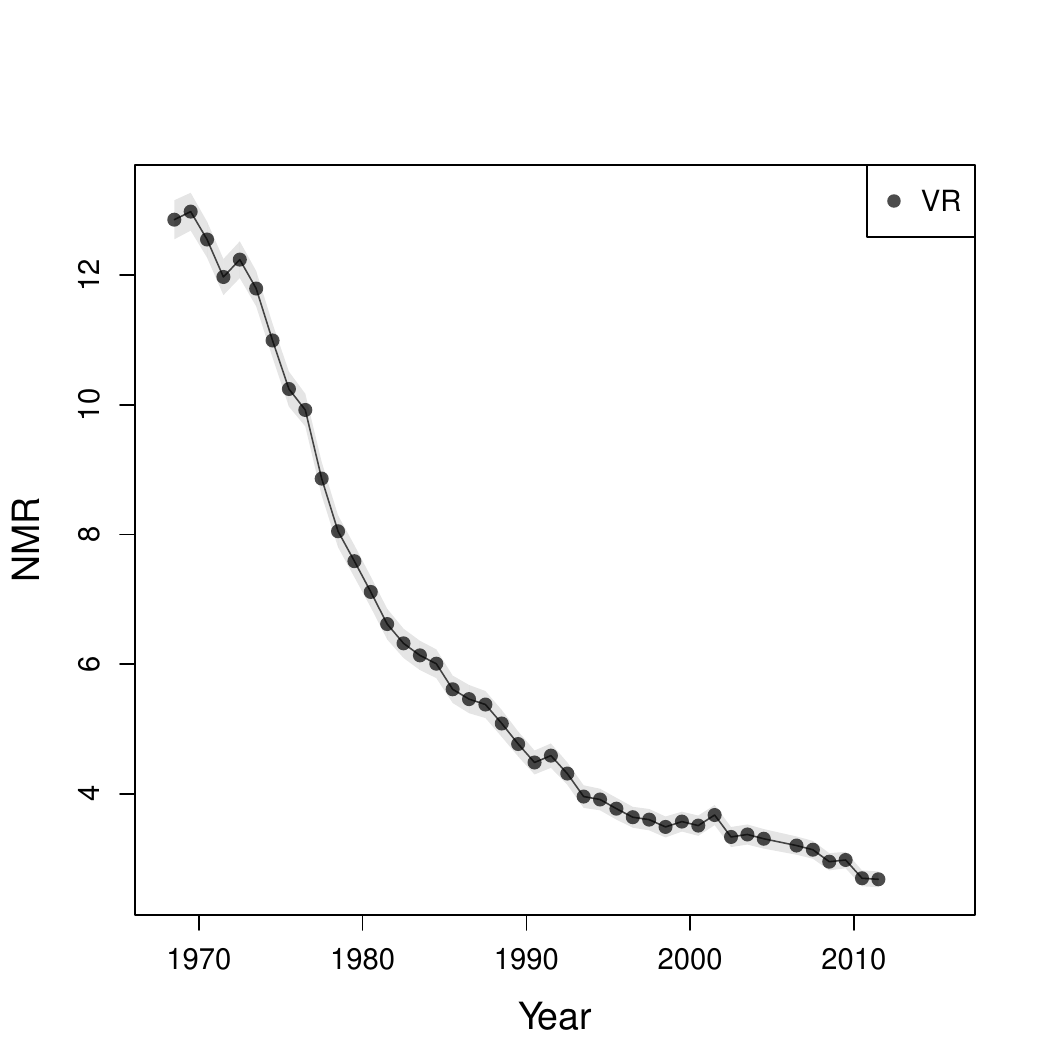} \caption{Australia} \label{fig:australia} 
\end{subfigure} ~ 
\quad 
\begin{subfigure}[b]{0.47\textwidth}  \includegraphics[width=\textwidth,  page=105]{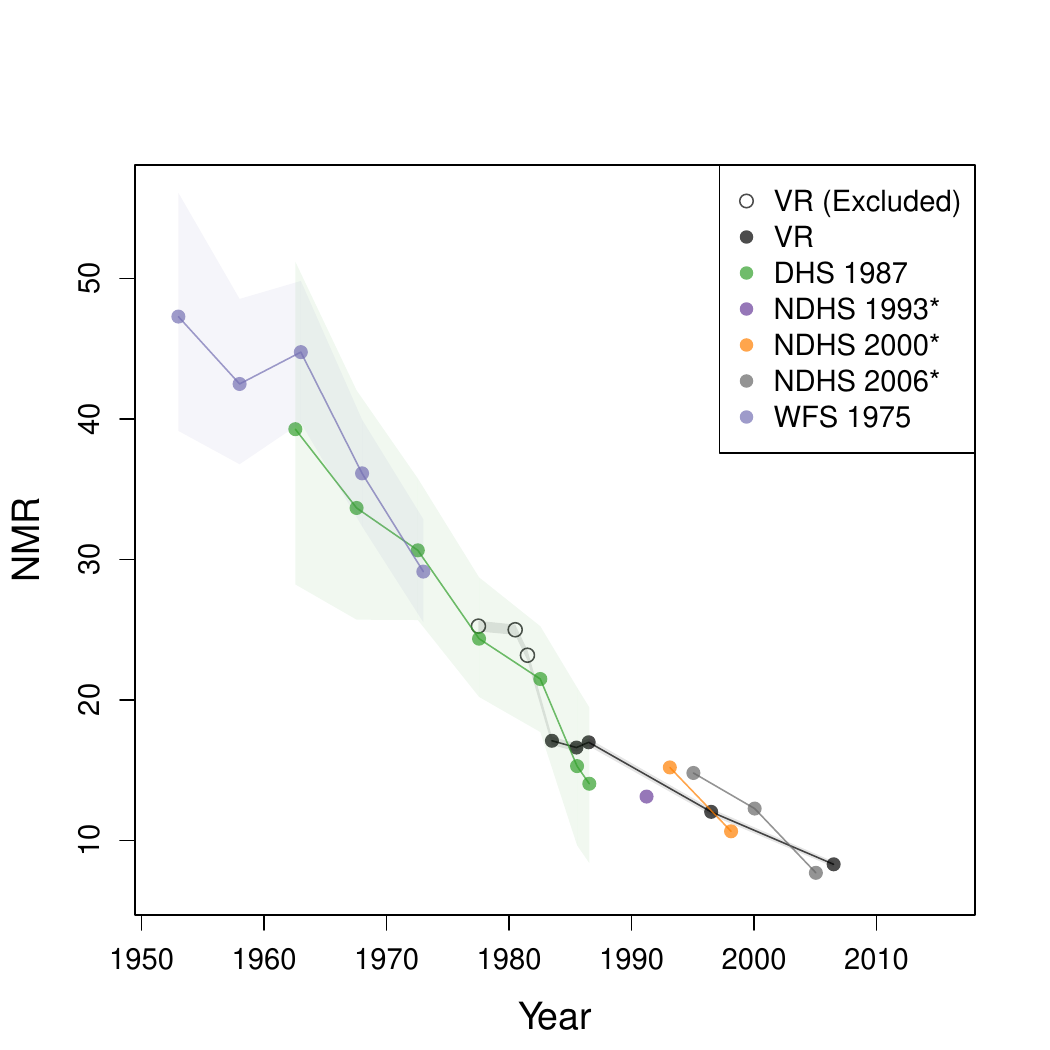} \caption{Sri Lanka} \label{fig:srilanka} 
\end{subfigure} ~ 
\\ 
\begin{subfigure}[b]{0.47\textwidth}   \includegraphics[width=\textwidth, page=56]{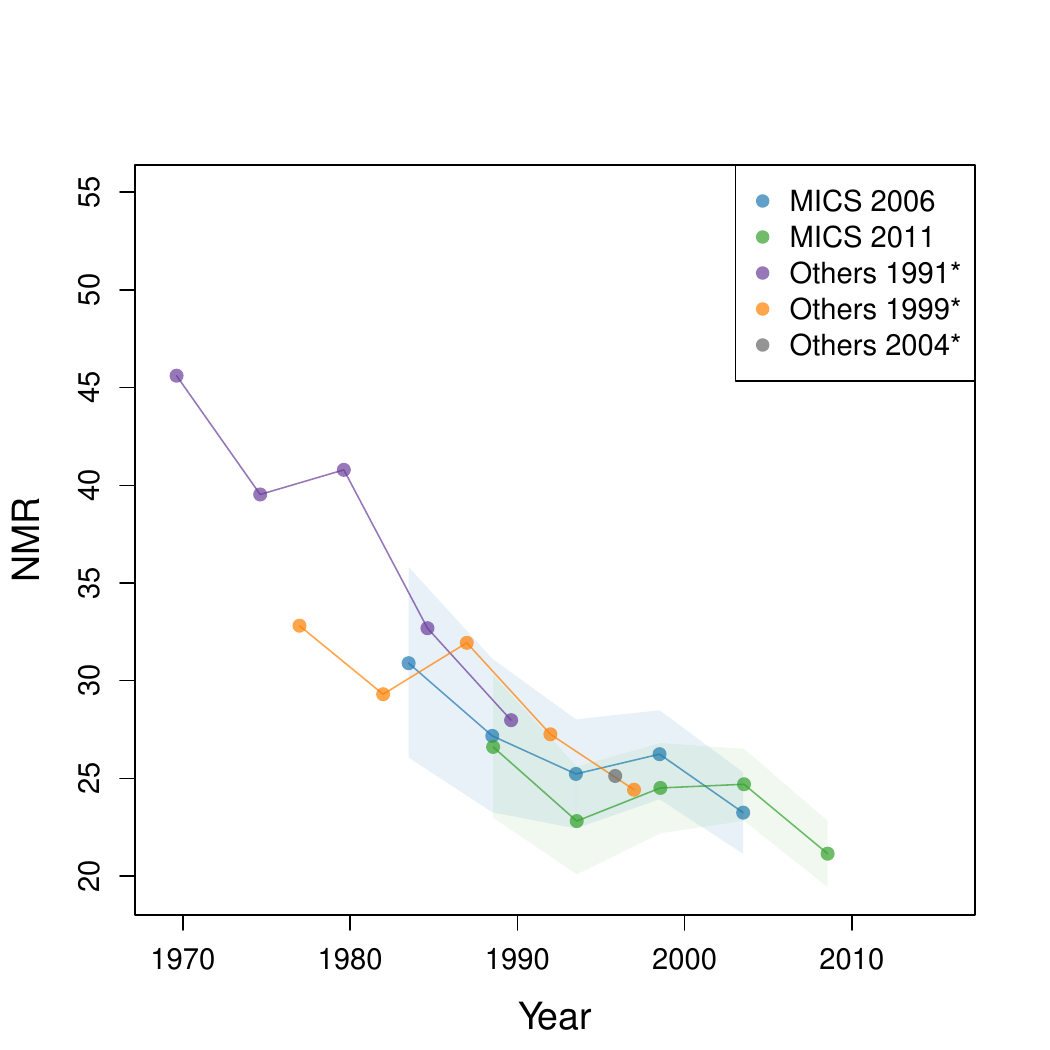} \caption{Iraq} \label{fig:iraq}
\end{subfigure} ~ 
\quad 
\begin{subfigure}[b]{0.47\textwidth}  \includegraphics[width=\textwidth, page= 125]{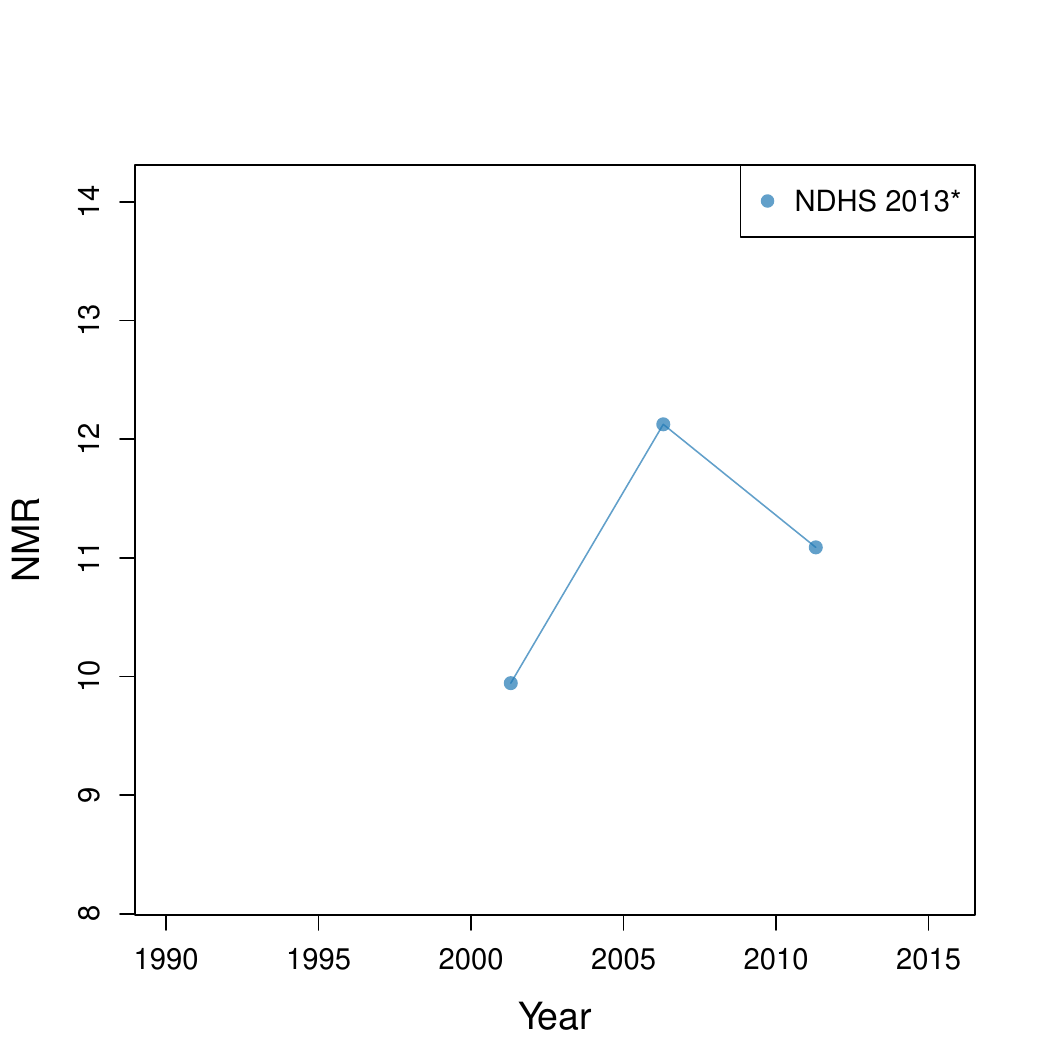} \caption{Vanuatu} \label{fig:vanuatu} 
\end{subfigure} ~ 
\caption{Neonatal Mortality Data (deaths per 1,000 births) for selected countries. Data series where sampling errors were not reported are marked with a *.}\label{data} 
\end{figure}

Figure \ref{data} illustrates examples of the data available for four countries. The shaded area around the observations has a width of two times the sampling or stochastic error. The NMR for Australia (Figure \ref{fig:australia}), as estimated from a full VR data time series, has a trend over time which is relatively regular and the uncertainty is low. Data for Sri Lanka indicate NMR are roughly five times as high as Australia. Data are available from 1950, but the VR data series is incomplete. The rest of the data comes from WFS, DHS and National DHS.  There are multiple estimates for some years, and the uncertainty around the estimates varies by source and year. The uncertainty around the VR data is much less than for the survey data. The National DHS series do not have estimates of sampling error. Iraq (Figure \ref{fig:iraq}) has no VR data, and the estimates are constructed from MICS and two other surveys: the Infant and Child Mortality and Nutrition Survey, and the Child and Maternal Mortality Survey. Again, there are multiple estimates for some time points, and uncertainty level and availability varies. Finally, Vanuatu (Figure \ref{fig:vanuatu}) has only three observation points from one National DHS. 

In the NMR model, country-year specific U5MRs were used as explanatory variables and to obtain final estimates.  Point estimates, given by posterior medians, as well as posterior samples were obtained from the UN IGME (IGME 2015).

\section{Method}
The aim is to produce estimates of the NMR for all countries in the world, and report the associated uncertainty around these estimates. The model needs to be flexible enough to estimate NMR in a variety of situations, as illustrated in Figure~\ref{data}. The estimates should follow the data closely for countries with reliable data and low uncertainty. On the other hand, the model estimates need to be adequately smooth in countries with relatively large uncertainty and erratic trajectories. The model also needs to be able to estimate NMR over the period 1990-2015 for \textit{all} countries, including those countries where there are limited or no data available. To do so, the model utilizes the relationship between the U5MR and NMR: as the level of U5MR decreases, the proportion of deaths under 5 that are neonatal increases. The model also allows for country-specific effects and time trends to capture data-driven trends in data-rich countries.

\subsection{Model overview}
Write $N_{c,t}$ and $U_{c,t}$ as the NMR and U5MR for country $c$ at time $t$, respectively, with $U_{c,t}$ given by the IGME U5MR estimate for that country-year.  We explain the model set-up in terms of the ratio 
\begin{equation*}
R_{c,t} = \frac{N_{c,t} }{U_{c,t} - N_{c,t}}, 
\end{equation*}
which refers to the (true) ratio of neonatal deaths compared to deaths in months 2 to 59. We constrain $R_{c,t}>0$ such that $0 \leq \frac{N_{c,t} }{U_{c,t} } \leq 1$ to guarantee that NMR estimates are not greater than U5MR estimates. The true ratio $R_{c,t}$ is modeled as follows:
\begin{equation}
R_{c,t} = f(U_{c,t})  \cdot P_{c,t}, \label{oriscale}
\end{equation}
where $f(U_{c,t})$ is the overall expected ratio given the current level of U5MR and $P_{c,t}$ is a country-specific multiplier to capture deviations from the overall relationship.

The observed ratio $r_{c,i}$, which refers to the $i$-th observation of the ratio in country $c$,  is expressed as a combination of the true ratio and some error, i.e.\
\begin{eqnarray}
\label{overallmodel}
r_{c,i} &=& R_{c,t[c,i]} \cdot \epsilon_{c,i}\\\nonumber
\implies \log(r_{c,i}) &=& \log(R_{c,t[c,i]}) + \delta_{c,i}
\end{eqnarray}
for $c=1,2, \hdots , C$ and $i=1,\hdots , n_{c}$, where $C=195$ (the total number of countries) and $n_{c}$ is the number of observations for country $c$. The index $t[c,i]$ refers to the observation year for the $i$-th observation in country $c$, $\epsilon_{c,i}$ is the error of observation $i$ and $\delta_{c,i} = \log(\varepsilon_{c,i})$.

\subsection{Modeling the ratio of neonatal to non-neonatal deaths}
\subsubsection{Global relationship with U5MR}
\begin{figure}[htbp]
\centering
\includegraphics[width=0.7\textwidth]{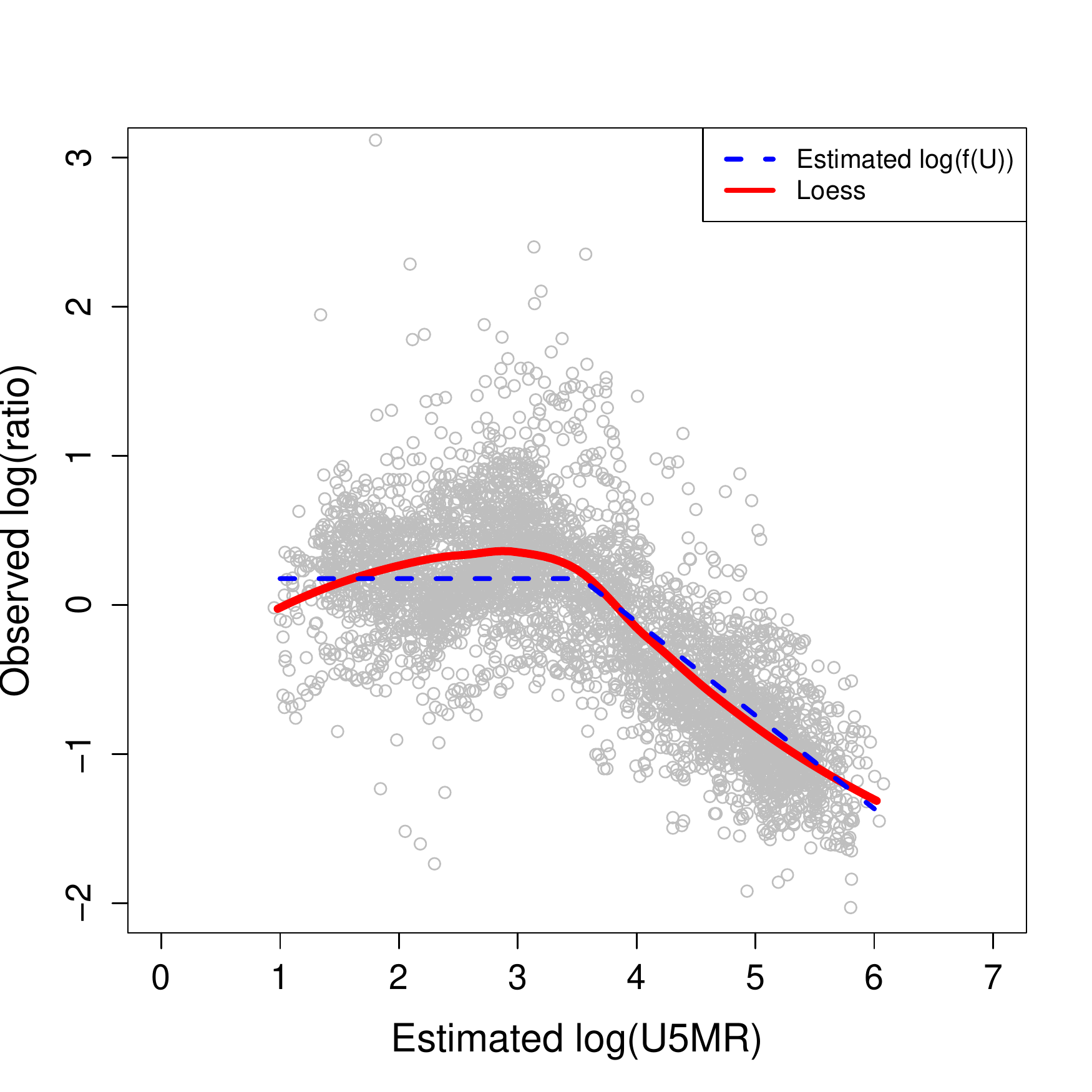}
\caption{Observed and estimated relation between the ratio of neonatal and non-neonatal deaths and under-five mortality. Observations $r_{c,i}$ are displayed with grey dots and plotted against $U_{c, t[c,i]}$. The Loess fit to the observations is shown in red, and the estimated relation (function $f(U_{c,t})$) is added in blue (dashed line).}
\label{scatter}
\end{figure}

The first step in modeling the ratio of neonatal to non-neonatal deaths is to find an appropriate function $f(\cdot)$ in Equation~\ref{oriscale}, which captures the expected value of the ratio given the current level of U5MR. Figure~\ref{scatter} shows a scatter plot of log-transformed observed ratios $\log(r_{c,i})$ versus $\log(U_{c,t[c,i]})$. The relationship between the two variables appears to be relatively constant up to around $\log(U_{c,t})=3$, after which point the log ratio decreases linearly with decreasing $\log(U_{c,t})$. Given this observed relationship, $f(\cdot)$ is modeled as follows:
\begin{equation*}
f(U_{c,t})= \beta_0 + \beta_1 \cdot \left(\log(U_{c,t}) - \log(\theta)\right)\bm{1}_{[U_{c,t}> \theta]},
\end{equation*}
with indicator function $\bm{1}_{[ U_{c,t}> \theta]}=1$ for $ U_{c,t}> \theta$ and zero otherwise such that $f(U_{c,t}) = \beta_0$ for $ U_{c,t}\leq  \theta$ and $f(U_{c,t})$ changes linearly with $\log(U_{c,t})$ with slope $\beta_1$ for $\log U_{c,t}>\log \theta$. Parameters $\beta_0$, $\beta_1$, and $\theta$ are unknown and are estimated. 

The fitted relationship between the ratio and the level of U5MR is illustrated in Figure~\ref{scatter}. The cutpoint $\theta$ is estimated to be around 34 deaths per 1,000 births (95\% CI: [33, 57]). At U5MR levels that are higher than $\theta$, the $\beta_1$ coefficient suggests that a 1\% increase in the U5MR leads to a 0.65\% decrease (95\% CI: [0.61, 0.71]) in the ratio $R_{c,t}$. The fitted line is quite similar in shape to the loess curve fitted to the data, shown by the red line in Figure~\ref{scatter}.

\subsubsection{Country-specific multiplier}
\begin{figure}[htbp] \centering 

\begin{subfigure}[b]{0.57\textwidth}  \includegraphics[width=\textwidth]{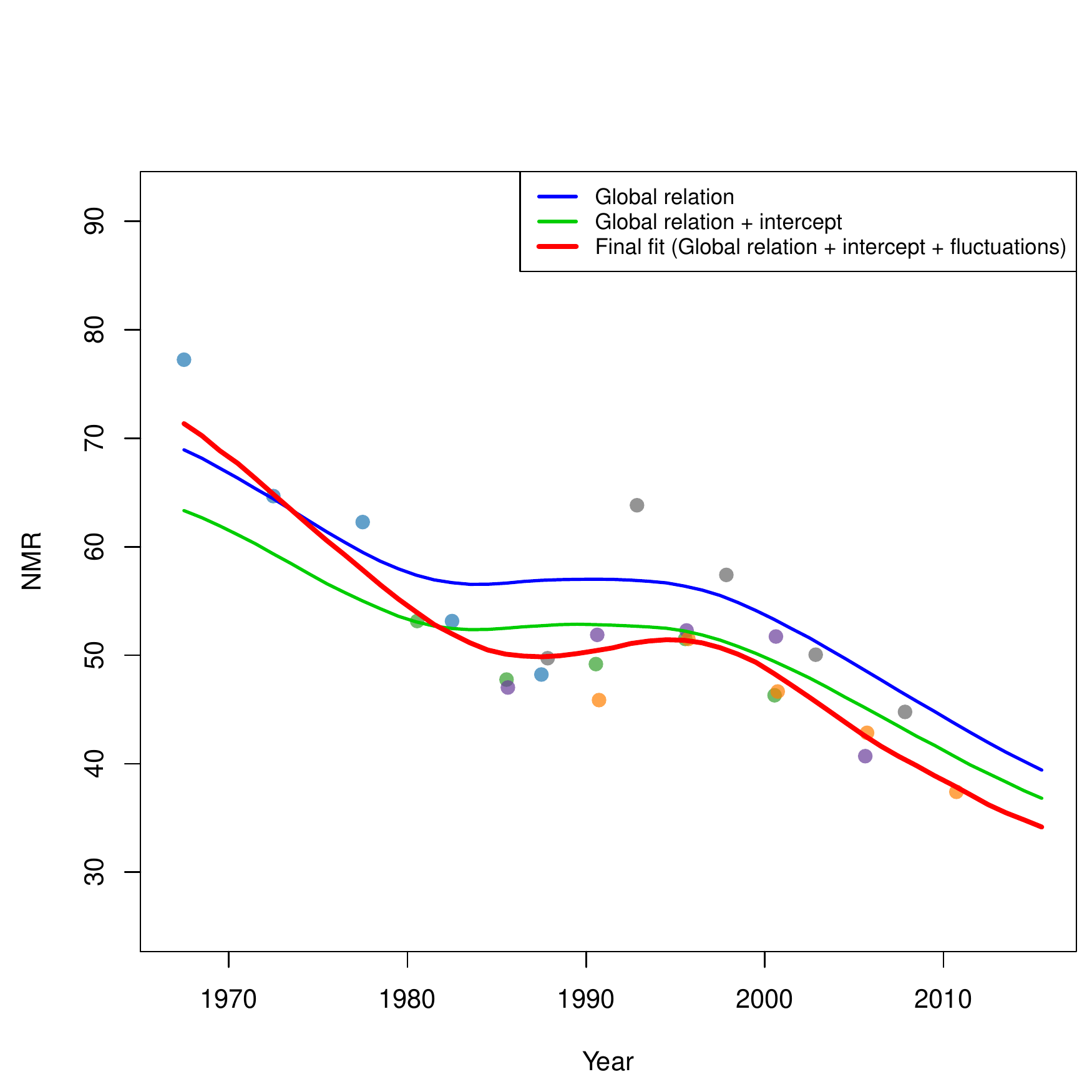} \caption{The three components of fit: the global relation $f(.)$, and $P_{c,t}$, which consists of a country-specifc intercept and fluctuations.} \label{threefits}  
\end{subfigure} ~ 
\\ 
\begin{subfigure}[b]{0.57\textwidth}  \includegraphics[width=\textwidth]{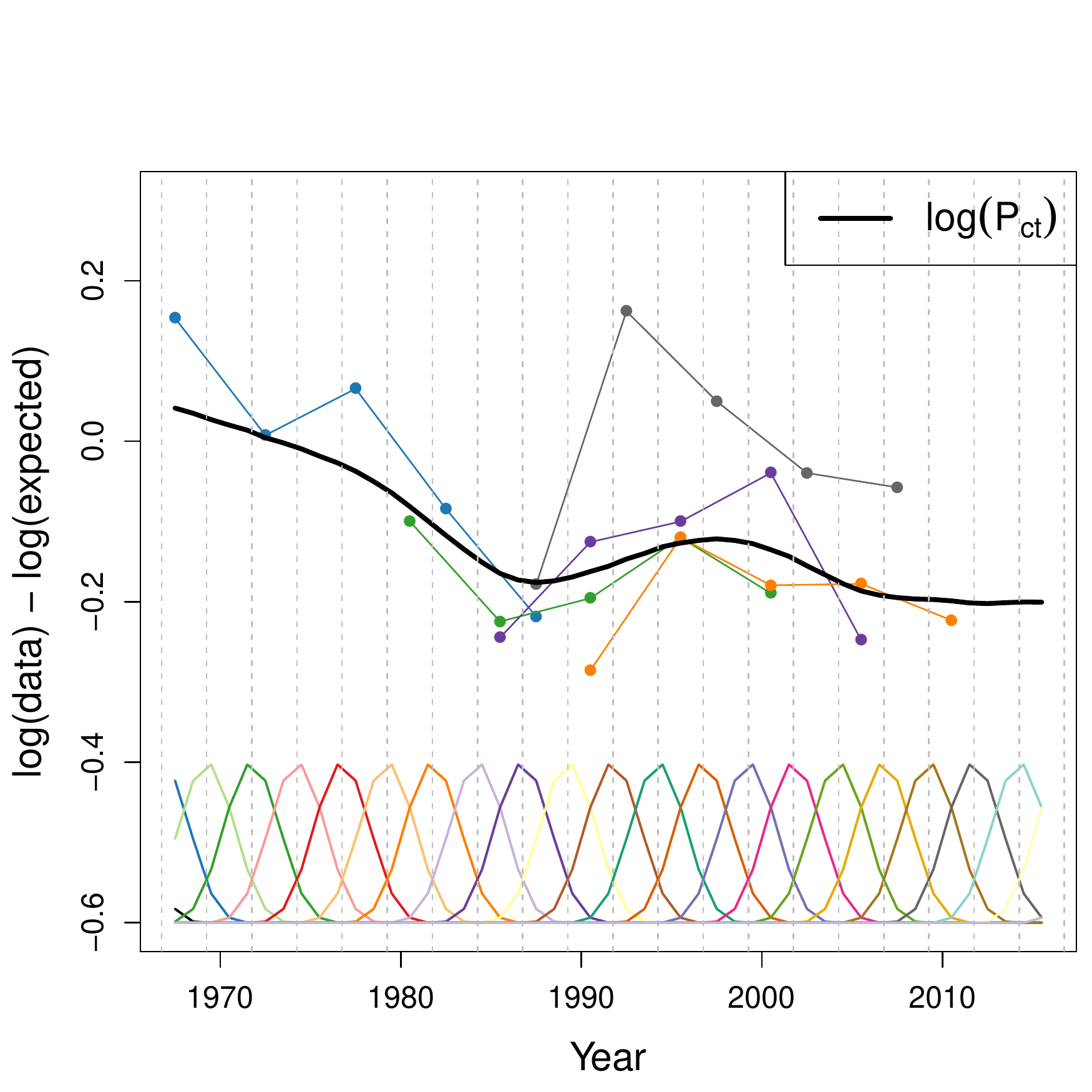} \caption{Estimate of $\log P_{c,t}$ for Nigeria using splines regression. The B-splines have been scaled vertically for display purposes. The gray dotted vertical lines indicate knot positions (every 2.5 years).  }  \label{NGAsplines} 
\end{subfigure} ~ 
\caption{Illustration of the global relation, splines regression and combined components for Nigera}\label{NGA} 
\end{figure}

Although there is a relationship between the neonatal ratio and U5MR at the aggregate level, the relationship  between $R_{c,t}$ and $U_{c,t}$ is likely to differ from country to country. For instance, some countries may have higher levels of NMR than what we expect given the level of U5MR. The model needs to account for this higher ratio of NMR to U5MR for these countries. However, within a particular country, the relationship between NMR and U5MR may not be constant over time, so the model should be flexible enough to also allow for temporal changes. This is the purpose of the country-specific term $P_{c,t}$ in Equation~\ref{oriscale}: to capture data-driven differences across countries and also within countries over time. The effect of the addition of $P_{c,t}$ for Nigeria is illustrated in Figure~\ref{threefits}. The blue dashed line represents the fit with just the the global relation $f(.)$, as described above. The inclusion of the $P_{c,t}$ changes the fit from the blue line to the red line, which allows the trajectory to follow the data more closely. 

The country-year multiplier $P_{c,t}$ was modeled on the log-scale with a B-splines regression model:
\begin{equation*}
\log(P_{c,t})= \sum_{k=1}^{K_c} B_{c,k}(t)\alpha_{c,k},
\end{equation*}
where $B_{c,k}(t)$ refers to the $k$th B-spline function for country $c$ evaluated at time $t$ and $\alpha_{c,k}$ is the $k$-th splines coefficient for country $c$.  The B-splines $B_{c,k}(t)$ were constructed using cubic splines. Country $c$ has a total of $K_c$ splines and $K_c$ knot points defined by $t_1<t_2< \dots <t_{K_c}$;  $K_c$ is the number of B-splines needed to cover the period up to 2015 and back to 1990 or the start of the observation period, whichever is earlier. In terms of knot spacing, the same interval length of 2.5 years was used in each country regardless of the number or spacing of observations. The consistent interval length was chosen to be able to exchange information across countries about the variability in changes between spline coefficients. 

The country-specific multiplier $P_{c,t}$ for Nigeria is illustrated in Figure \ref{NGAsplines}. Each spline is represented in a different color at the bottom of the figure, and the gray dotted vertical lines indicate knot positions. The splines regression is fit to residual pattern in the data on the log scale once the global relation is taken into account. 

A penalty was imposed on the first-order differences between spline coefficients $\alpha_{c,k}$ to ensure a relatively smooth fit. To better see how the smoothing penalty is implemented, we rewrite the $k$-th splines coefficient for country $c$, $\alpha_{c,k}$, in the form of an intercept and fluctuations around the intercept (Eilers and Marx 1996; Currie and Durban, 2002):
\begin{equation}
\label{alphas}
\alpha_{c,k} = \lambda_c + [\boldsymbol{D}_{K_c}'(\boldsymbol{D}_{K_c}\boldsymbol{D}_{K_c}')^{-1}\boldsymbol{\varepsilon}_c]_k,
\end{equation}
where $\boldsymbol{D}_{K_c}$ is a first-order difference matrix: ${D}_{K_c i,i} = -1$, ${D}_{K_c i,i+1}=1$ and ${D}_{K_c i,j} = 0$ otherwise; and $\boldsymbol{\varepsilon}_c$ is a vector of length $Q_c$:
\begin{equation*}
  \boldsymbol{\varepsilon}_c = (\varepsilon_{c,1}...\varepsilon_{c, Q_c})',
\end{equation*}  
where $Q_c = K_c-1$. Each element of the vector is the first order difference of the coefficients $\alpha_{c,q}$ i.e.\
\begin{equation*}
 \varepsilon_{c,q} = \alpha_{c,q+1} - \alpha_{c,q} = \Delta\alpha_{c, q}
\end{equation*} 
  for $q=1...Q_c$. 

The $\lambda_c$ represent country-specific deviations in level from the overall global relationship between $R_{c,t}$ and $U_{c,t}$. As such the $\lambda_c$'s are modeled centered at zero so that
\begin{equation*}
\lambda_c \sim N(0, \sigma_{\lambda}^2).
\end{equation*}

The $\lambda_c$ is similar to a country-specific intercept. If $\lambda_c>1$, then country $c$'s level of NMR is generally higher than expected given its U5MR level, and vice versa for $\lambda_c<1$. For Nigeria (Figure~\ref{threefits}), the estimated country specific intercept is negative, and so the addition of the country-specific intercept lowers the fit from the blue line (global relation only) to the green line.

The $\boldsymbol{\varepsilon}_c$ term represents fluctuations around this country-specific intercept in the form of first-order differences in adjacent spline coefficients. These fluctuation terms allow for the $P_{c,t}$ term to be influenced by the changes in the level of the underlying data. The first-order differences in adjacent spline coefficients are penalized to guarantee the smoothness of the resulting trajectory as follows:
\begin{equation*}
\varepsilon_{c,q} \sim N(0, \sigma_{\varepsilon_c}^2),
\end{equation*}
where variance $\sigma_{\varepsilon_c}^2$ essentially acts as a country-specific smoothing parameter. The smoothness of a particular country's trajectory depends on the regularity of the trend in the data and also the measurement errors associated with the data points. As $\sigma_{\varepsilon_c}^2$ decreases, the fluctuations go to zero, and the $\alpha_{c,k}$'s become a country-specific intercept with no change over time. The $\sigma_{\varepsilon_c}^2$  is modeled hierarchically:
\begin{equation}
\label{smoothingequation}
\log(\sigma_{\varepsilon_c}^2) \sim N(\chi, \psi_{\sigma}^2),
\end{equation}
where $e^{\chi}$ can be interpreted as a `global smoothing parameter' and $\psi_{\sigma}^2$ reflects the across-country variability in smoothing parameters. The hierarchical structure of the model allows information on the amount of smoothing to be shared across countries.  The countries with fewer data points and thus less information about the level of smoothness borrow strength from countries with more observations. 

\clearpage 
\subsection{Data model}
Equation~\ref{overallmodel} indicates the observed ratio $r_{c,i}$ is modeled on the log-scale as the true ratio ${R_{c,t}}$ plus some error term $\delta_{c,i}$. This error term $\delta_{c,i}$ is modeled differently based on the source of the data. 

\paragraph{VR data}\label{VRdatamodel}

For VR data series, the error term $\delta_{c,i}$ is modeled as 
\begin{equation*}
\delta_{c,i} \sim N(0,  \tau_{c,i}^2 ), 
\end{equation*}
where  $\tau_{c,i}^2$ is the stochastic standard error. These are obtained based on standard assumptions about the distribution of deaths in the first month of life. Details are given in the Appendix. 

\paragraph{Non-VR data}
For the non-VR data, the error term $\delta_i$ is modeled as 
\begin{equation*}
\delta_{c,i} \sim N(0,  \nu_{c,i}^2 + \omega^2_{s[c,i]}), 
\end{equation*}
where $\nu_{c,i}$ is the sampling error and $\omega_{s[c,i]}$ is non-sampling error of the series type $s$ of observation $i$ in country $c$. Non-sampling error variances are estimated separately for each of the series types listed in Table~\ref{tbledata}: DHS, Other DHS, MICS, and Others. The distinction by series type was made to allow for the possibility that a particular survey group may run a survey in a similar fashion across all countries, and as such may display similar characteristics in terms of non-sampling error.  

Sampling error variances were reported for the majority of the non-VR observations (see Table~\ref{tbledata}). For those observations $(c,i)$ where sampling error was not reported, the sampling error was imputed based on the median value of all observed sampling errors of series type $s[c,i]$ within the group-size category of country $c$, which is small or other. A country was categorized as `small' if the annual number of births was in the lowest quartile of all countries (corresponding to a maximum of around 25,000 births per year). The distinction between small and other countries was made due to the large differences in observed standard errors. The imputed values for missing standard errors for each size category and series type are shown in Table \ref{SEs}.

\begin{table}[htbp]
\centering
\caption{Values imputed for missing standard errors for survey data by series and country size category}
\label{SEs}
\begin{tabular}{l|c|c}
&\multicolumn{2}{|c}{Country size category}\\\hline
Series type          & Other & Small \\ \hline
DHS       & 0.13      & 0.26  \\
MICS      & 0.16      & 0.21 \\
Other DHS & 0.14      & 0.24  \\
Others    & 0.16      & 0.22 
\end{tabular}
\end{table}

\subsection{Obtaining the final estimates} \label{finalest}
The model described so far produces estimates of $R_{c,t}$. The corresponding estimate of $N_{c,t}$ is obtained by transforming the ratio and combining it with $U_{c,t}$:
\begin{equation*}
N_{c,t} =\mbox{logit}^{-1} (R_{c,t}) \cdot U_{c,t}.
\end{equation*}
To take into account the level of uncertainty in the $U_{c,t}$ estimates, the $N_{c,t}$ estimates are generated by randomly combining posterior draws of $R_{c,t}$ and of $U_{c,t}$. The result is a series of trajectories of $N_{c,t}$ over time. The best estimate is taken to be the median of these trajectories and the 2.5th and 97.5th percentiles are used to construct 95\% credible intervals.

\subsection{Other aspects of the method}
Other aspects of the method, including the projection method, estimation for countries with no data and crisis and HIV/AIDS adjustments are detailed in Appendix B.

\subsection{Computation} 
The hierarchical model detailed in the previous sections is summarized in Appendix \ref{modelsumm}. The model was fitted in a Bayesian framework using the statistical software R. Samples were taken from the posterior distributions of the parameters via a Markov Chain Monte Carlo (MCMC) algorithm. This was performed through the use of JAGS software (Plummer 2003). 

In terms of computation, three chains with different starting points were run with a total of 20,000 iterations in each chain. Of these, the first 10,000 iterations in each chain were discarded as burn-in and every 10th iteration after was retained. Thus 1,000 samples were retained from each chain, meaning there were 3,000 samples retained for each estimated parameter. 

Trace plots were checked to ensure adequate mixing and that the chains were past the burn-in phase. Gelman's $\hat{R}$ (Gelman and Rubin 1992) and the effective sample size were checked to ensure a large enough and representative sample from the
posterior distribution. The value of $\hat{R}$ for all parameters estimated was less than 1.1. 

\section{Results}
Estimates of NMR were produced for the 195 UN member countries for at least the period 1990 -- 2015, with periods starting earlier if data were available. In this section some key results are highlighted. Results are also compared to those produced by the method previously used by the IGME. 

The estimated global relation (Table~\ref{globalcoeff}) suggests that the relationship between the ratio and U5MR is constant up to a U5MR of 34 deaths per 1,000 births, the ratio of neonatal to other child mortality is constant at around 1.20 (95\% CI: [1.03, 1.25]). This is equivalent to saying the proportion of deaths under-five that are neonatal is constant at around 54 per cent (95\% CI: [50, 55]). Above a U5MR 34/1000, the estimated coefficient suggests that, at the global level, a 1 per cent increase in the U5MR leads to a 0.6 per cent decrease in the ratio.

\begin{table}[h!]
\centering
\caption{Estimates for parameters in global relation}
\label{globalcoeff}
\begin{tabular}{lrr}
          &  Median     & 95\% CI       \\ \hline
$\beta_0$ & 0.18  & (0.03, 0.22)   \\
$\beta_1$ & -0.62 & (-0.61, -0.71)  \\
$U_{cut}$ & 34.27 & (33.73, 57.34)\\ \hline
\end{tabular}
\end{table}

\subsection{Results for selected countries}
\begin{figure} \centering 

\begin{subfigure}[b]{0.47\textwidth} \includegraphics[width=\textwidth, page=130]{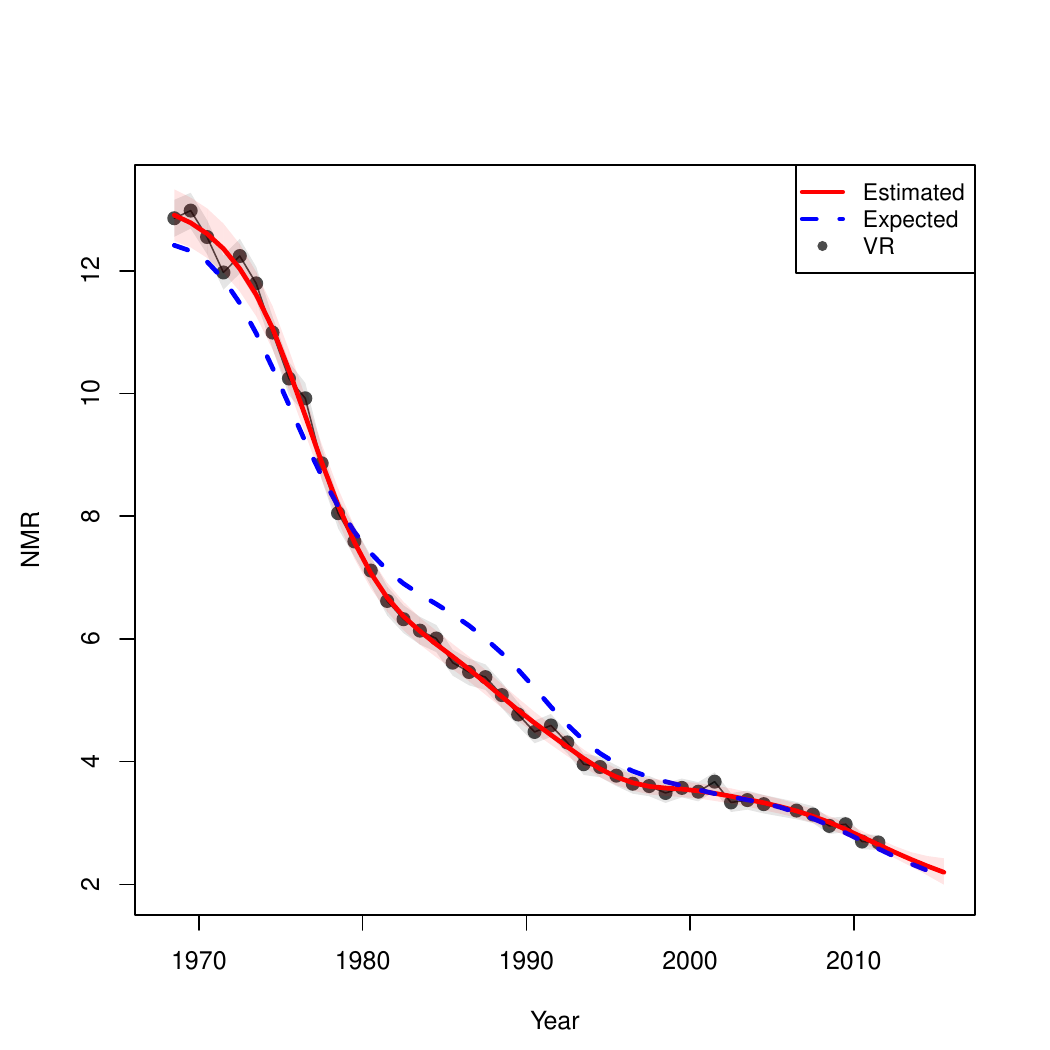} \caption{Australia} \label{fig:australiafit}  
\end{subfigure} ~ 
\quad 
\begin{subfigure}[b]{0.47\textwidth}  \includegraphics[width=\textwidth, page=105]{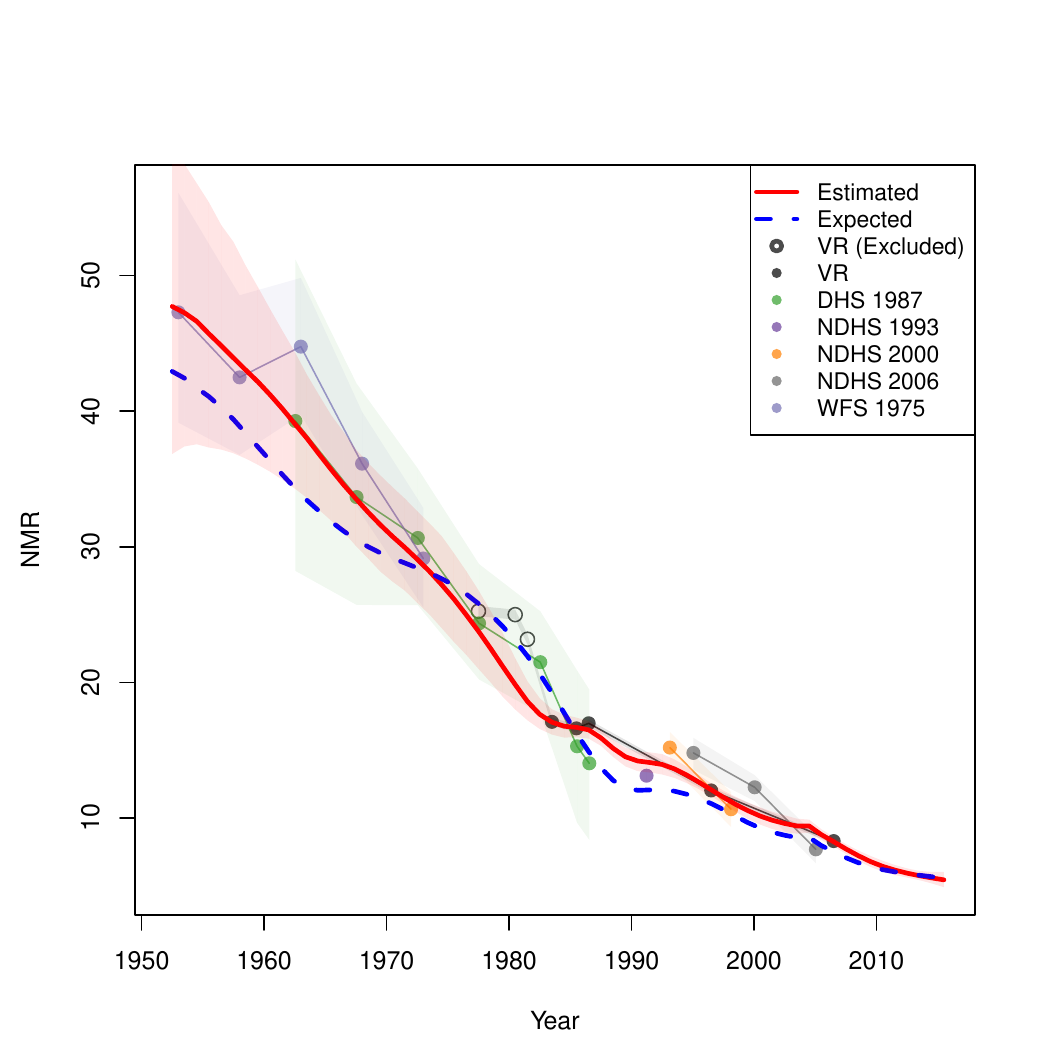} \caption{Sri Lanka} \label{fig:srilankafit} 
\end{subfigure} ~ 
\\ 
\begin{subfigure}[b]{0.47\textwidth} \includegraphics[width=\textwidth, page=56]{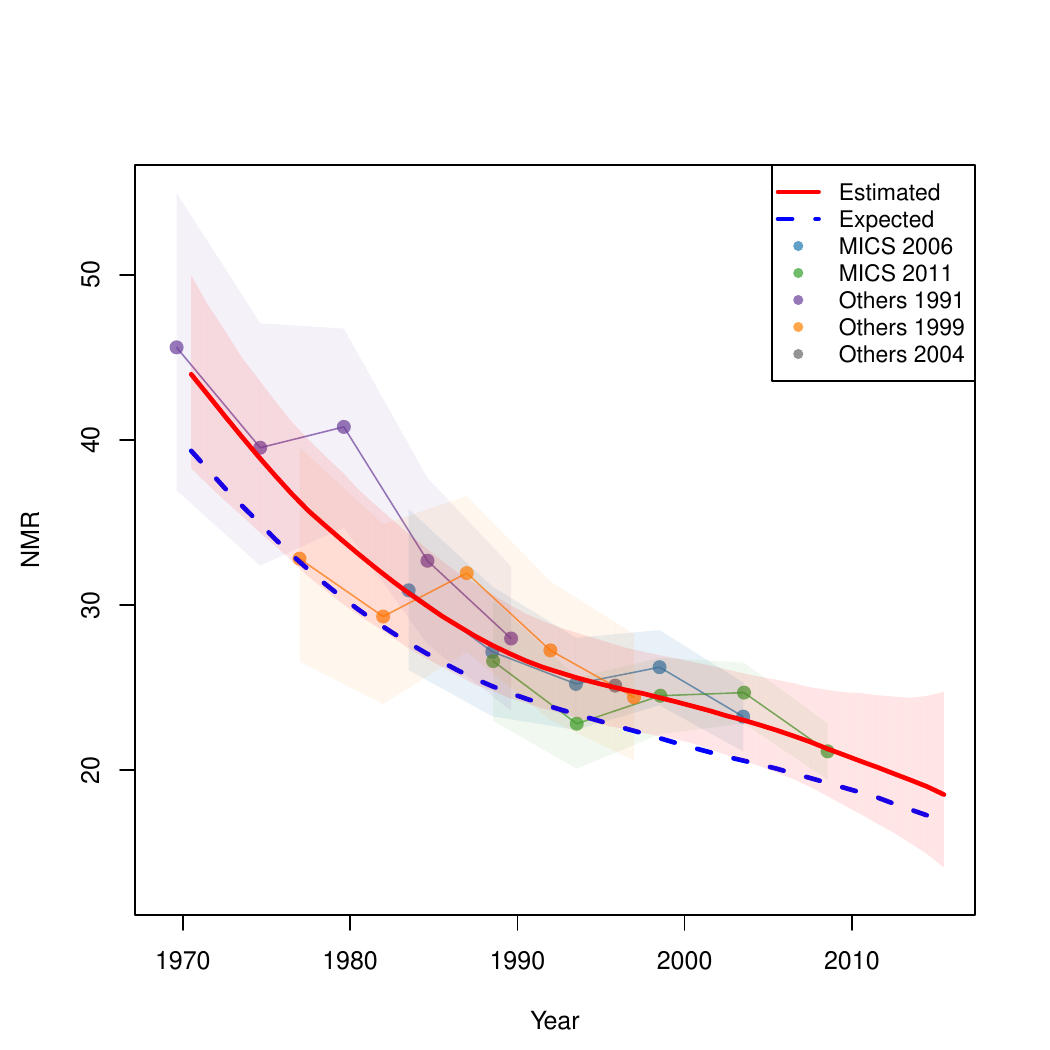}  \caption{Iraq} \label{fig:iraqfit} 
\end{subfigure} ~ 
\quad 
\begin{subfigure}[b]{0.47\textwidth}  \includegraphics[width=\textwidth, page=125]{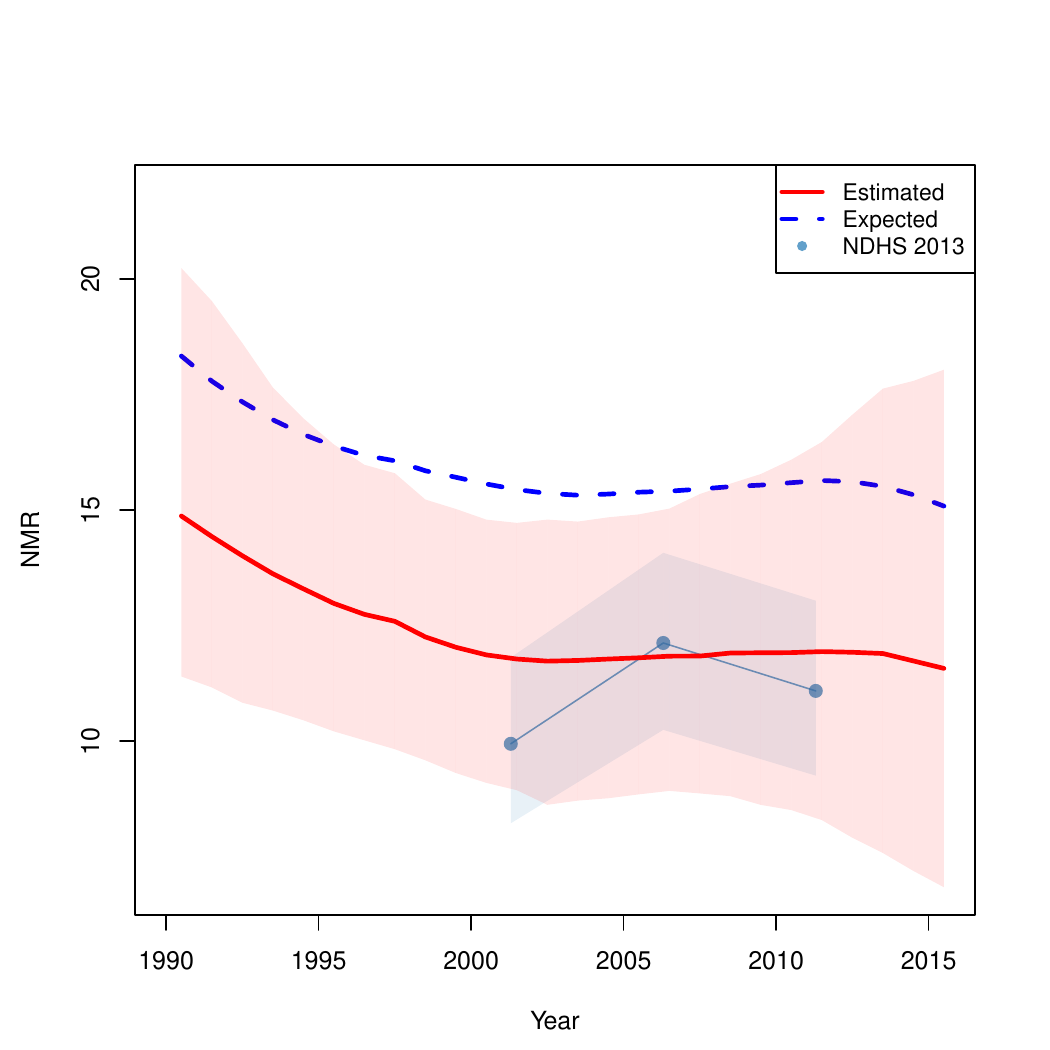} \caption{Vanuatu} \label{fig:vanuatufit} 
\end{subfigure} ~ 
\caption{Observed and estimated neonatal mortality (deaths per 1,000 births) for selected countries}\label{Results} 
\end{figure}

The fits for the four countries illustrated in Section 2 are shown in in Figures \ref{Results}. For Australia  (Figure \ref{fig:australiafit}), the estimated red line follows the data closely, given the small uncertainty levels around the data. There has been a steady decrease in NMR since 1970. In earlier time periods, the level of NMR was higher than the expected level (that is, the red line is higher than the blue). This switched in the 1980s and 1990s, and more recently, the estimated and expected levels are close. 

For Sri Lanka (Figure \ref{fig:srilankafit}), the estimates of NMR are informed by the combination of VR and survey data. The VR has a greater influence on the trajectory because of the smaller associated standard errors. In the earlier years, the uncertainty intervals around the estimate are larger due to the higher uncertainty of the data. There is a small spike in the estimate in the year 2004, which is a tsunami-related crisis adjustment. 

No VR data were available for Iraq (Figure \ref{fig:iraqfit}), and the larger sampling errors around the survey data have led to relatively wide uncertainty intervals over the entire period. This is in contrast to Sri Lanka, where uncertainty intervals became more narrow once VR data was available. The larger sampling errors in Iraq have also led to a relatively smooth fit (high smoothing parameter), and the shape of the trajectory essentially follows the shape of the expected line. 

For Vanuatu (Figure \ref{fig:vanuatufit}), the trajectory is driven by the expected trajectory given Vanuatu's trend in U5MR. The available data determine the country-specific intercept for Vanuatu, which is lower than the expected level. However, the relative absence of data for this country means that the uncertainty around the estimates is high.

\subsection{Outlying countries}

\begin{figure} \centering 
\includegraphics[width=0.7\textwidth]{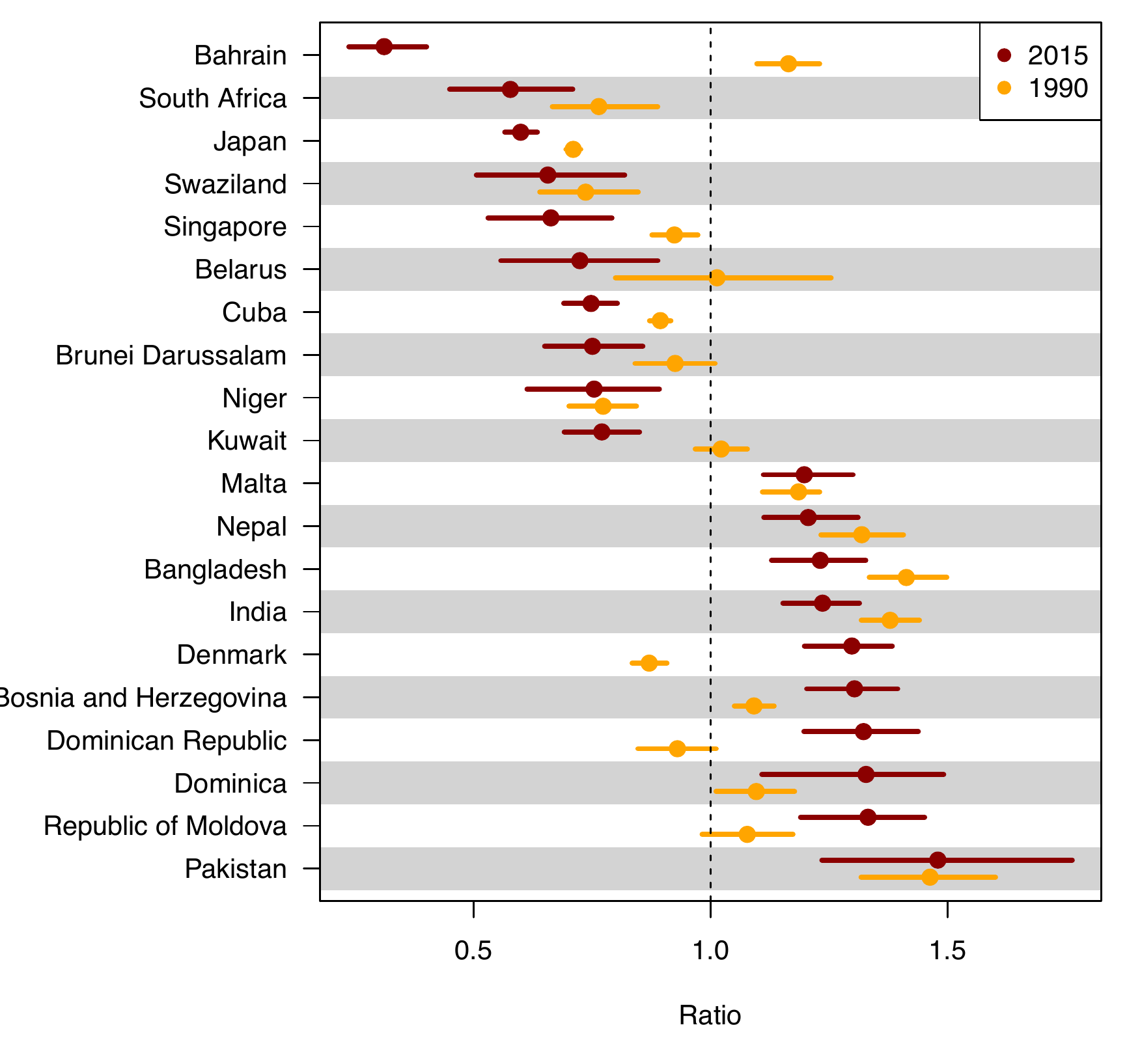}
\caption{Ratio of estimated to expected NMR for outlying countries}\label{ratios} 
\end{figure}

The set-up of the model allows for the intuitive interpretation of results in comparing the estimated level of NMR to the expected level given the U5MR. We define a country to be outlying if the the estimated NMR in 2015 was significantly higher or lower than the expected level by at least 10 per cent. That is, the ratio of estimated-to-expected was at least 1.1 or less than 0.9 in 2015, and statistically significant at the 5 per cent level. Figure~\ref{ratios} illustrates these countries, and the values of estimated-to-expected in 1990 and 2015. 

Countries that have a lower-than-expected NMR include Japan, Singapore and South Korea, and some African countries such as South Africa and Swaziland. Countries that have a higher-than-expected NMR include several Southern Asia countries, such as Bangladesh, Nepal, India and Pakistan. The former Yugoslavian countries Croatia, Bosnia and Herzegovina, and Montenegro also have higher-than-expected NMR. 

\begin{figure} \centering 
\begin{subfigure}[b]{0.47\textwidth}  \includegraphics[width=\textwidth, page=150]{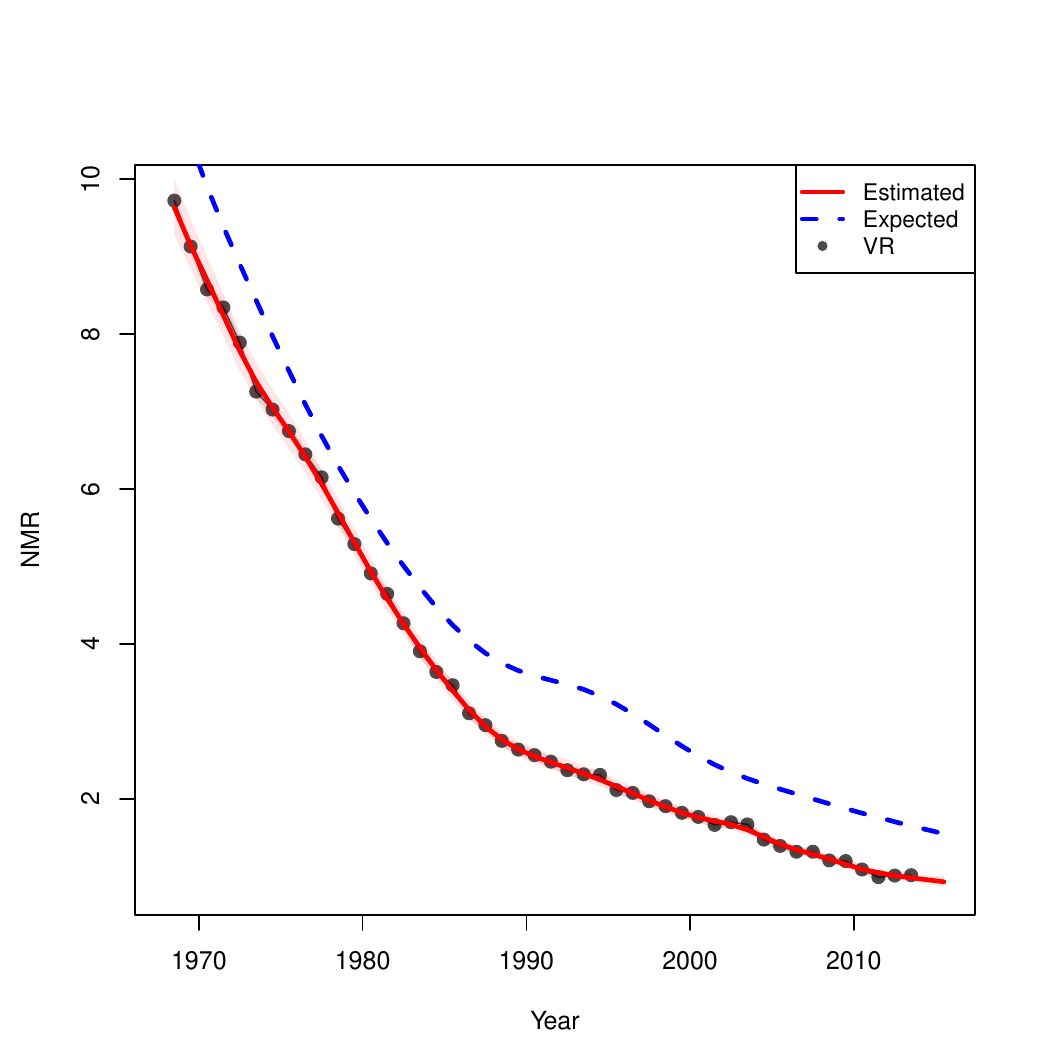} \caption{Japan} \label{japan} 
\end{subfigure} ~ 
\begin{subfigure}[b]{0.47\textwidth}  \includegraphics[width=\textwidth,page=53]{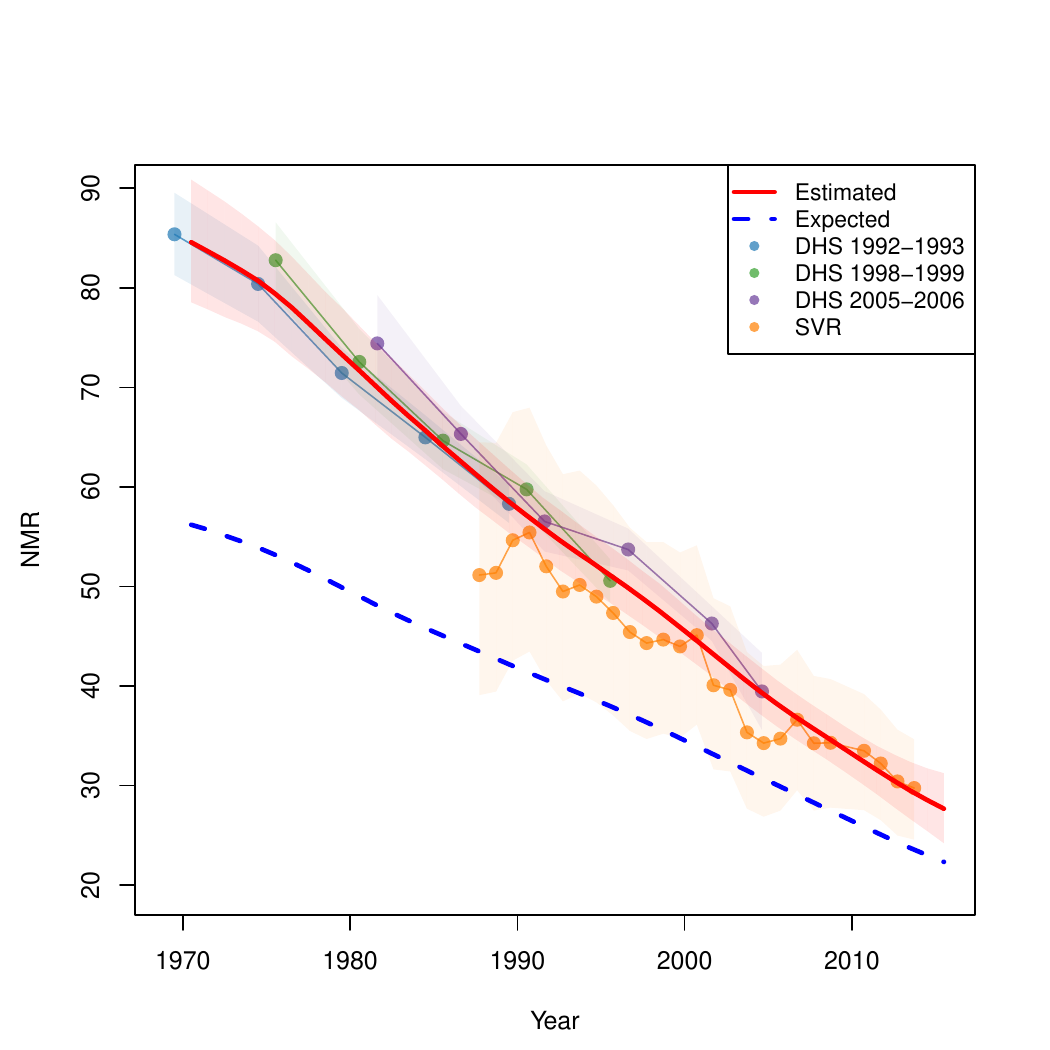}  \caption{India} \label{india} 
\end{subfigure} ~ 
\caption{Higher and lower-than-expected countries}\label{outlying} 
\end{figure}

Figure~\ref{outlying} shows estimates through time for two contrasting countries, Japan, which has lower-than-expected NMR and India, with higher-than-expected NMR. In each of the figures, the red line represents the estimated fitted line (with 90\% CIs). The blue line represents the estimation with the $f(U_{c,t})$ only (without the country-specific effect, $P_{c,t}$). The blue line can be interpreted as the expected level of NMR in a particular year given the level of U5MR. The gap between the expected and estimated is being sustained through time for Japan, and has widened since the 1970s. The change in NMR levels for India has been dramatic. Not only is the current NMR around 30 per cent of what it was in 1970, the discrepancy between the expected and estimated levels is decreasing through time, and is no longer significant.

\subsection{Smoothing}
\begin{figure}[htbp]
\centering
\includegraphics[width=0.7\textwidth]{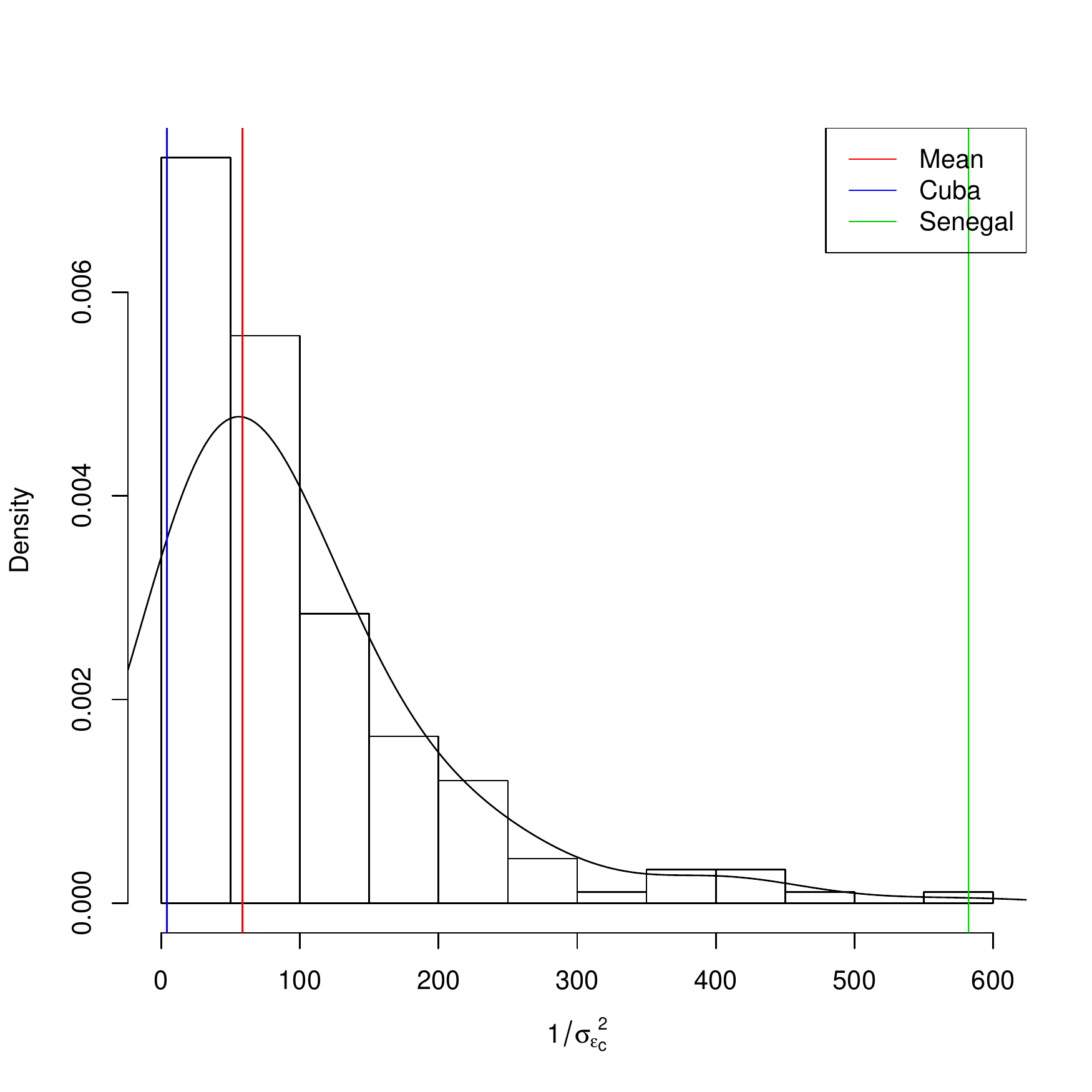}
\caption{}
\label{smoothinghist}
\end{figure}

The smoothness of the fluctuations $\sigma^2_{\varepsilon_c}$ is modeled hierarchically, assuming a log-normal distribution with a mean parameter $\chi$ (see Equation \ref{smoothingequation}). Smoothing parameters can also be expressed in terms of precision, $1/\sigma^2_{\varepsilon_c}$; Figure~\ref{smoothinghist} shows the distribution of estimated precisions for all countries. The larger the value of the smoothing parameter (precision), the smoother the fit. The estimate of the mean smoothing parameter was around 59 (95\% CI: [40, 83]). 

\begin{figure} \centering 
\begin{subfigure}[b]{0.47\textwidth}  \includegraphics[width=\textwidth, page=101]{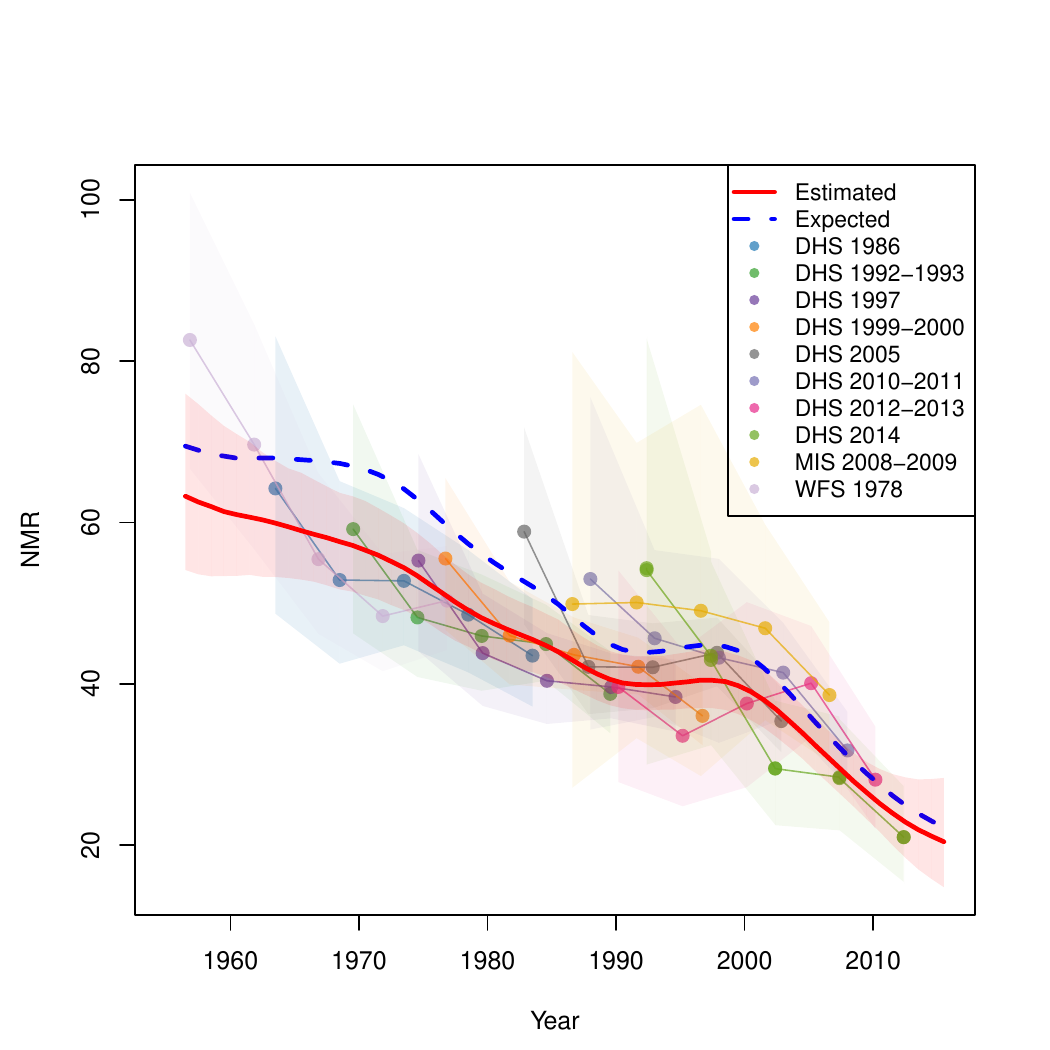} \caption{Senegal} \label{senegal} 
\end{subfigure} ~ 
\begin{subfigure}[b]{0.47\textwidth}  \includegraphics[width=\textwidth,page=33]{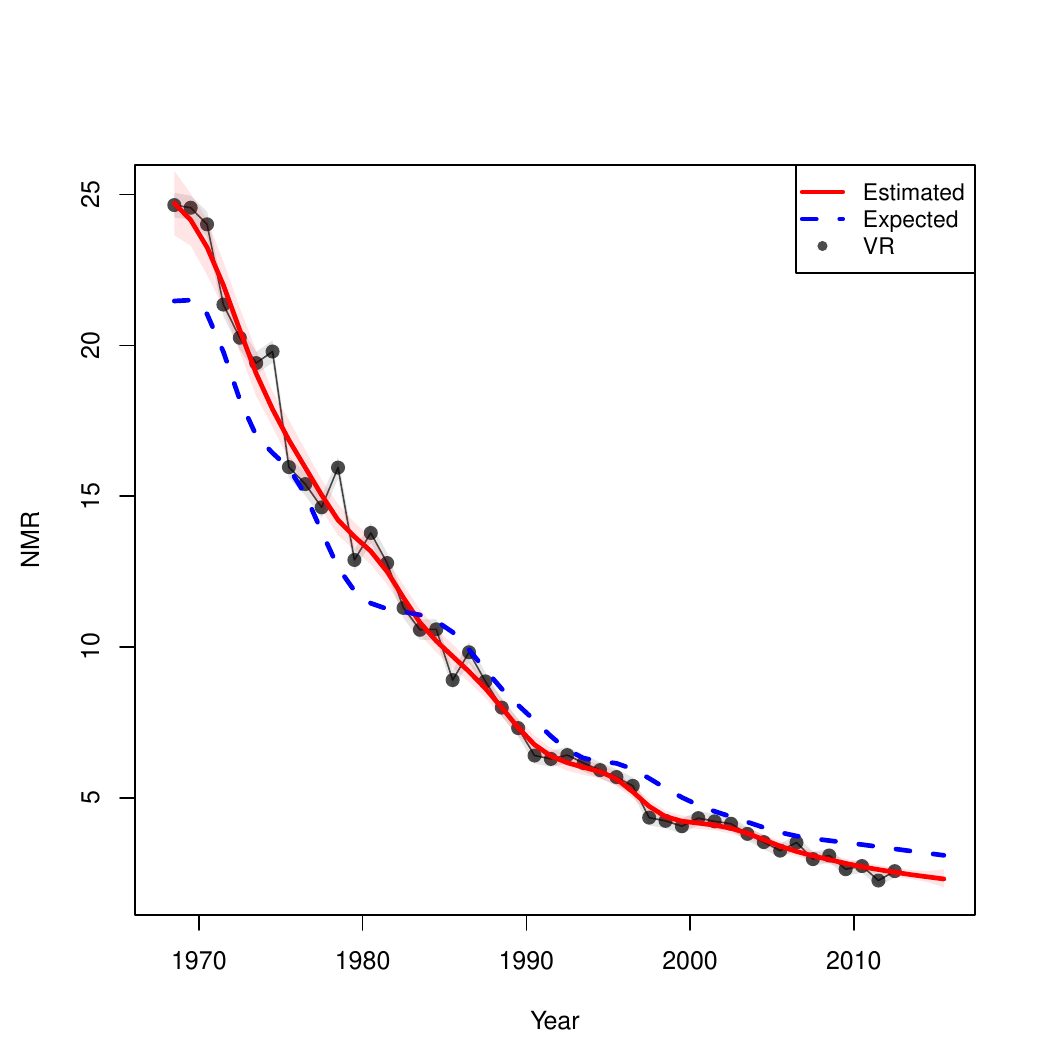}  \caption{Cuba} \label{cuba} 
\end{subfigure} ~ 
\caption{Countries with small and large amount of smoothing}\label{smoothingex} 
\end{figure}

Larger values of smoothing parameters were estimated for countries that had no available VR data but many observations from survey data. Senegal, which had the highest smoothing parameter at a value of 582 (95\% CI: [85, 5828]), had a total of 55 observations over a 45-year period (Figure~\ref{senegal}). The effect of having many observations with relatively large standard errors is a relatively smooth fit. In contrast, one of the smallest smoothing parameters occurred for Cuba, at around 4 (95\% CI: [2, 8]). Cuba is a country with good quality VR-data, which has relatively small standard errors. This means the fitted line follows the data more closely (Figure~\ref{cuba}).

\subsection{Comparison with existing IGME model}
It is useful to compare the results of this new model to the NMR results from the model previously used by the IGME. The previous model is described in Oestergaard et al.\ (2011). In this method, NMR estimates for countries with complete VR series are taken directly from the data. For countries without a complete VR series, a multilevel model is fit using U5MR as a predictor. A quadratic relationship with U5MR is specified. In addition, the model allows for country-level and region-level random effects:

\begin{equation*}
\log (NMR_{c,t}) = \underbrace{\alpha_0 + \beta_1 \log(U_{c,t}) + \beta_2  (\log U_{c,t})^2}_{\log \left(f(U_{c,t}) \right)} + \underbrace{\alpha_{country[i]} + \alpha_{region[i]} }_{\log \left( P_{c,t} \right)}
\end{equation*}

For comparison, the new model is:
\begin{eqnarray*}
\log (R_{c,t}) &=& \underbrace{\beta_0 + \beta_1 \cdot \left(\log(U_{c,t}) - \log(\theta)\right)_{[\log(U_{c,t})>\log(\theta)]}}_{\log \left(f(U_{c,t}) \right)} \\
&+& \underbrace{\sum_k^{K_c} B_{c,k}(t)\alpha_{c,k}}_{\log \left( P_{c,t} \right)}
\end{eqnarray*}

The existing model is similar in that it estimates NMR as a function of U5MR, plus some additional country-specific effect, i.e.\ there is an $f(U_{c,t})$ and a $P_{c,t}$. However, one of the main differences between the two models is that for countries with non-VR data, estimates from the new model can be driven by the data, while the previous model is restricted to follow the trajectory of the U5MR in a particular country, plus or minus some country-specific intercept. The other differences between the two models are highlighted in Table \ref{compare}.

\begin{table}[h!]
\centering
\caption{Comparison of two models}
\label{compare}
\begin{tabular}{c|c}
IGME 2014                                            & New model                                               \\ \hline
Model used for non-VR countries                      & Model used for all countries                            \\
$f(U_{c,t})$ is quadratic                            & $f(U_{c,t})$ is linear with changing slope              \\
$P_{c,t}$ is a country and region-specific intercept & $P_{c,t}$ is a country-specific int + fluctuations \\
Country-specific effect constant over time           & Country-specific effect can change over time            \\
No data model                                        & Data model                                             
\end{tabular}
\end{table}

\begin{figure} \centering 

\begin{subfigure}[b]{0.47\textwidth} \includegraphics[width=\textwidth, page=11]{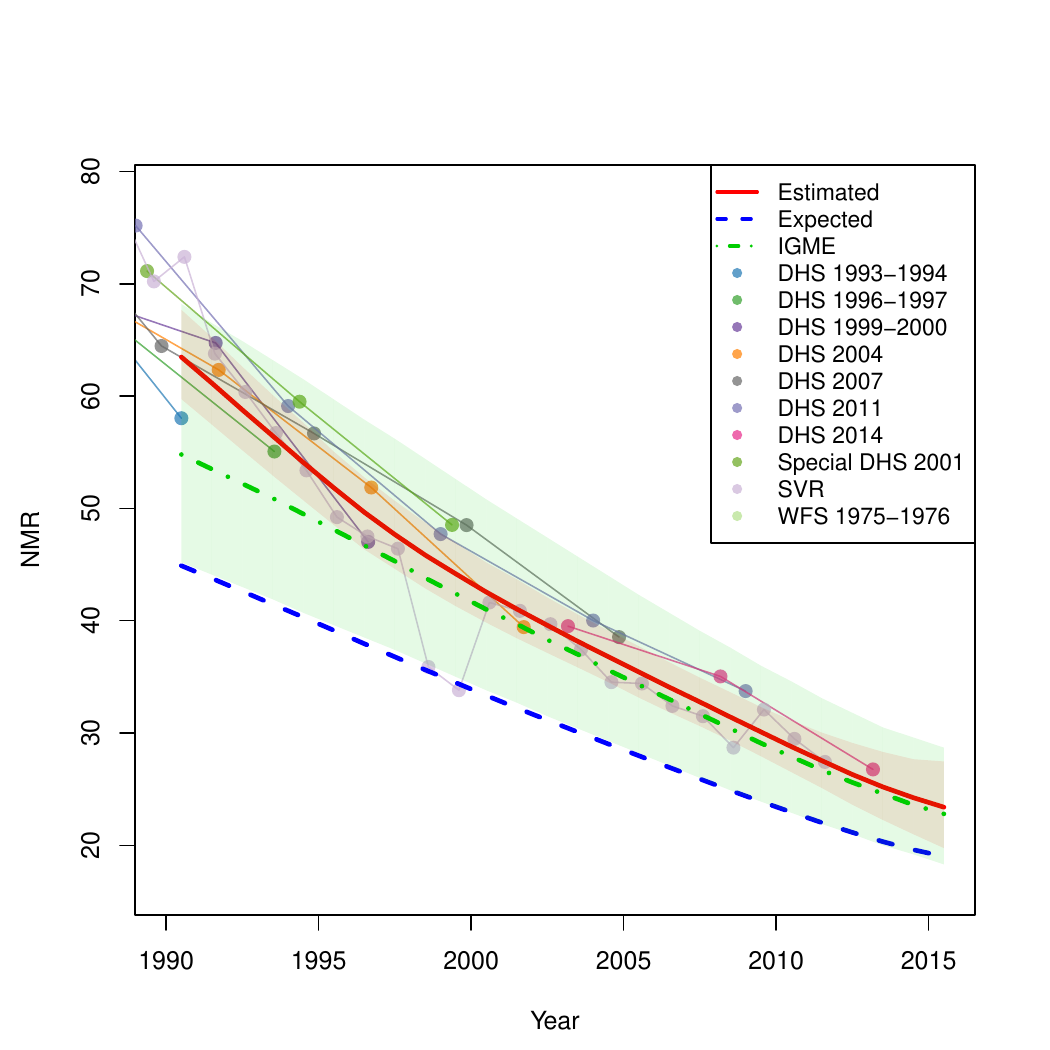} \caption{Bangladesh} \label{fig:BGDfit}  
\end{subfigure} ~ 
\quad 
\begin{subfigure}[b]{0.47\textwidth}   \includegraphics[width=\textwidth, page=21]{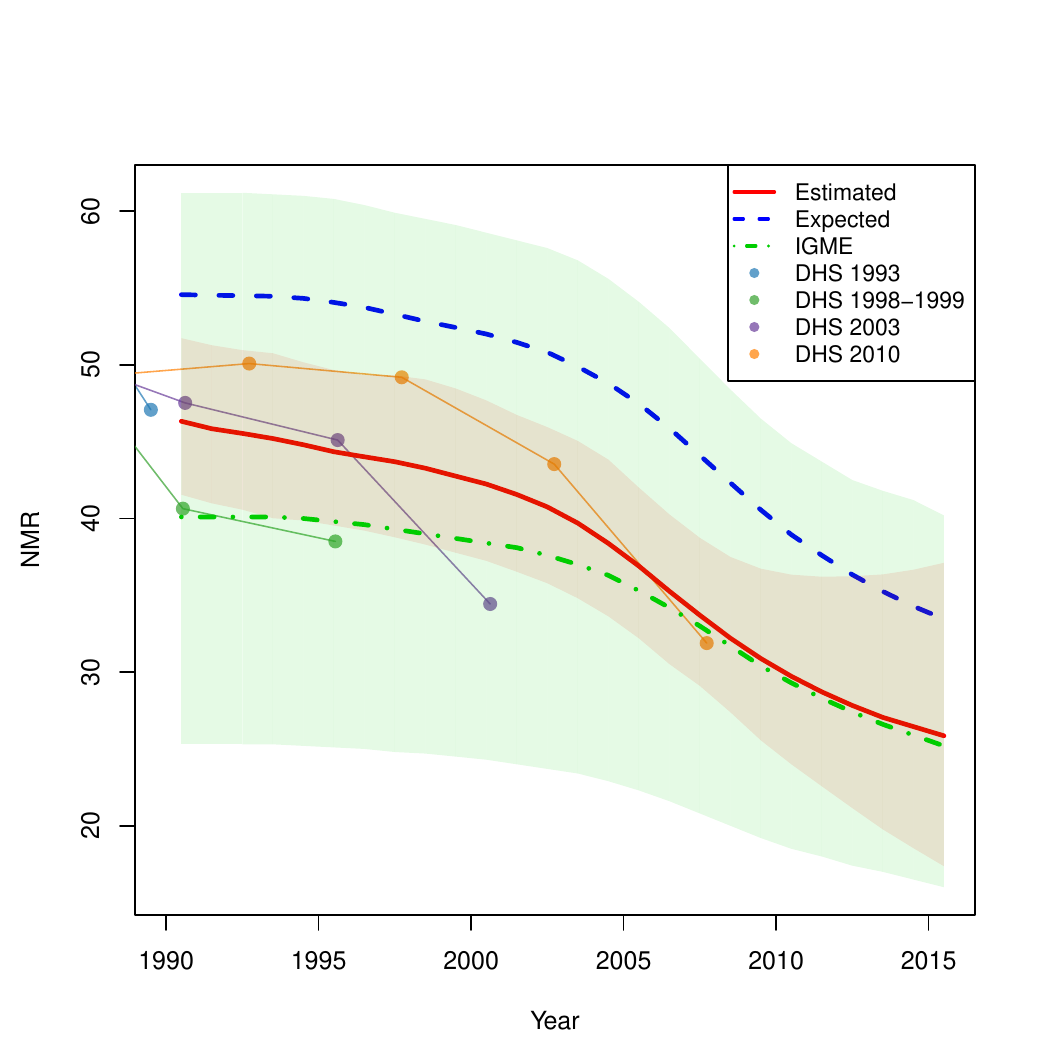} \caption{Burkina Faso} \label{fig:BFAfit}
 \end{subfigure} ~ 
\\ 
\begin{subfigure}[b]{0.47\textwidth}   \includegraphics[width=\textwidth, page=183]{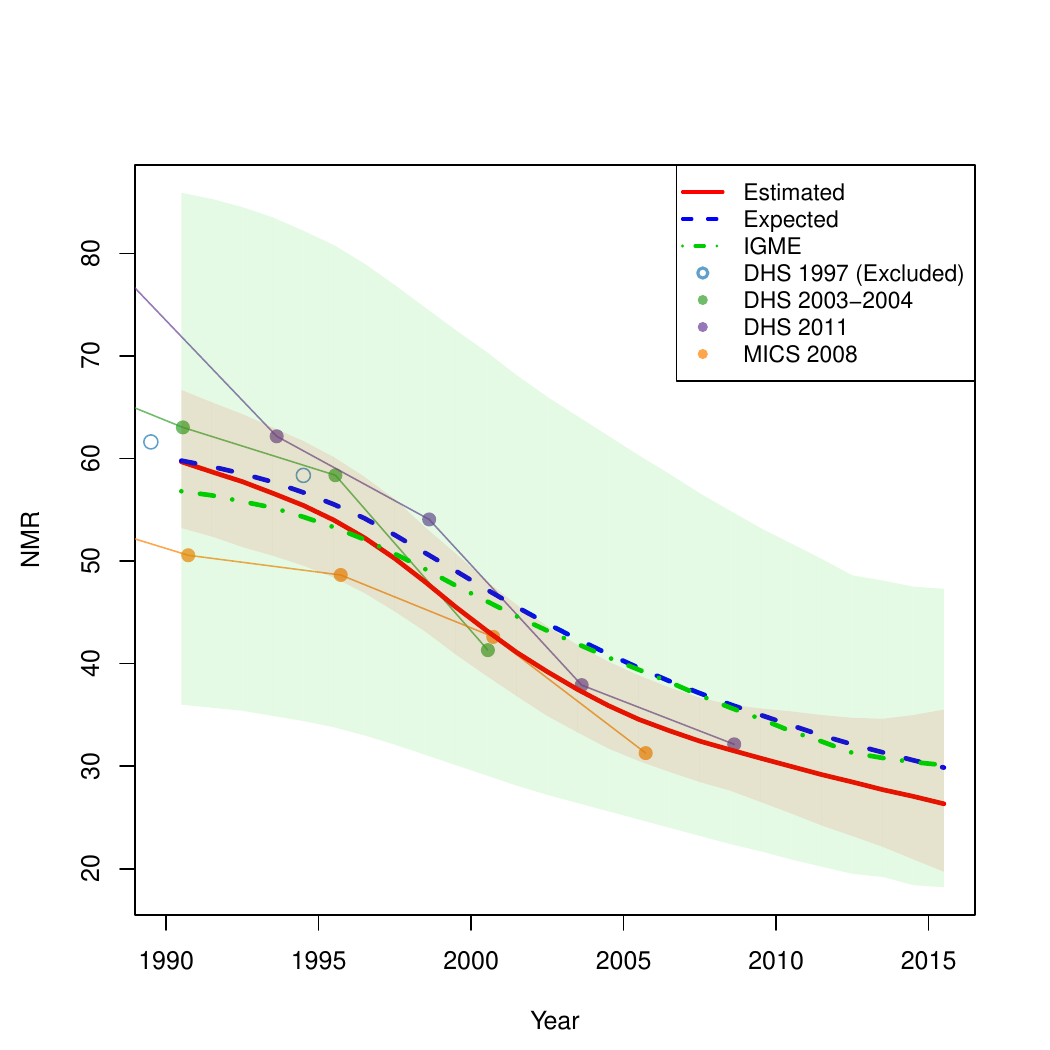} \caption{Mozambique} \label{fig:MOZfit}
\end{subfigure} ~ 
\quad 
\begin{subfigure}[b]{0.47\textwidth}  \includegraphics[width=\textwidth, page=35]{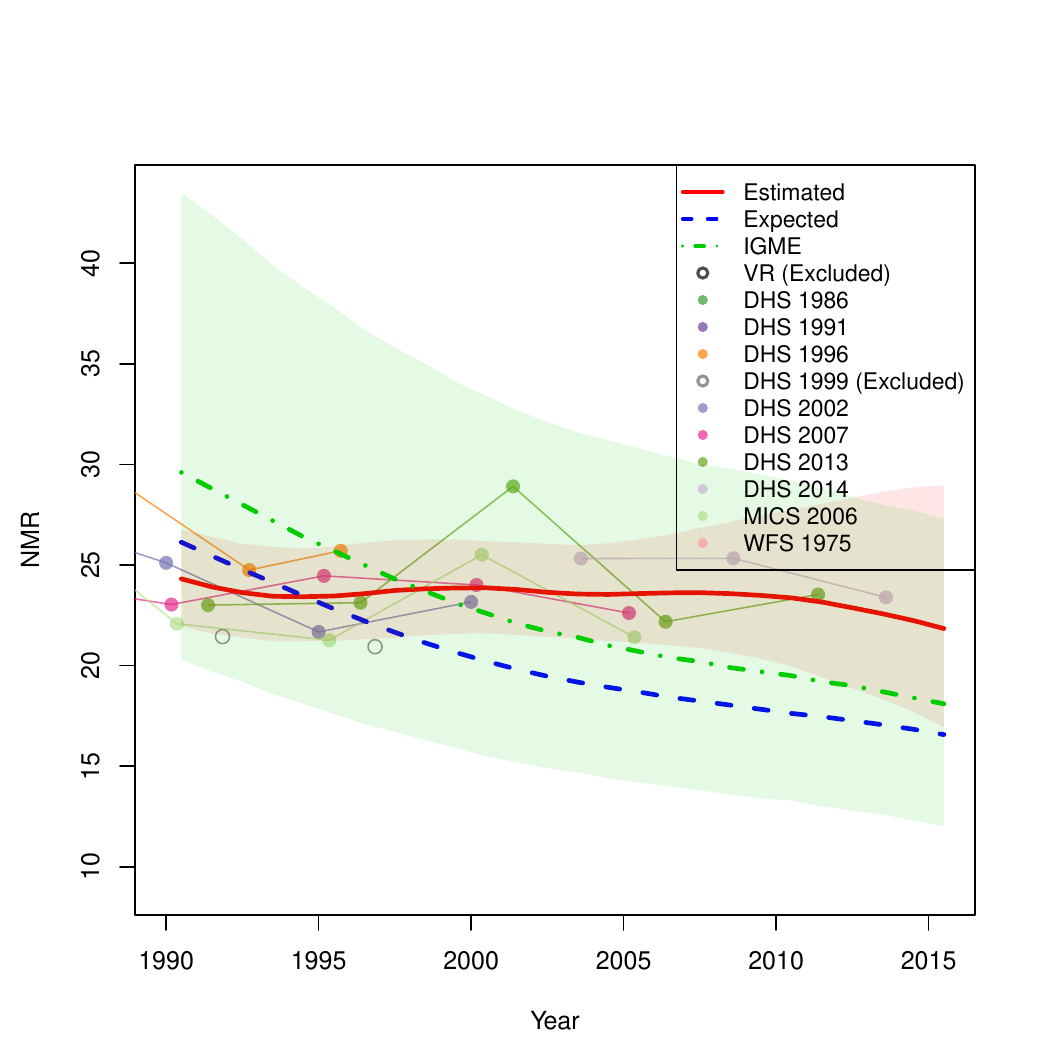} \caption{Dominican Republic} \label{fig:DOMfit} 
\end{subfigure} ~ 
\caption{Estimated Neonatal Mortality (deaths per 1,000 births); new versus IGME 2014 model}\label{ResultsI} 
\end{figure}

Figure \ref{ResultsI} compares the results of four countries to the estimates from the current IGME model. The charts are focused to just show the period 1990--2015, which is the period for which the IGME publishes estimates.The estimates from the existing IGME model generally follow the same trajectory as the expected line, as determined by U5MR patterns, and is shifted up or down depending on the estimate of the country-specific effect. In contrast, the estimates from the new model follow the data more closely. The fluctuation part of the country-specific multiplier $P_{c,t}$ allows the estimated line to move above or below the expected line, as is the case with the Dominican Republic (Figure \ref{fig:DOMfit}). In addition there is generally less uncertainty around the estimates in the new model, especially in periods where there are data. 

\subsection{Model validation} 
Model performance was assessed through an out-of-sample model validation exercise. In creating a training dataset, rather than removing observations at random, the process of removing data was chosen to emulate the way in which new data may be received. Mortality databases are updated as least once a year as more data become available. These updates may include not only data for the most recent time period but may also include, for example, retrospective estimates from a survey. Ideally the model should not be sensitive to updates of historical data so estimates do not change from year to year. 

The training set was constructed by leaving out the most recent survey data series, and for countries with only one series (including VR countries), the most recent 20 per cent of data observations were removed. The resulting training data set was made of around 80 per cent of the total data available. 

For the left-out observations, the absolute relative error is defined by 
\begin{equation*}
e_i =  \frac{| n_i - \tilde{n}_i |}{\tilde{n}_i},
\end{equation*}
where where $\tilde{n}_i$ denotes the posterior median of the predictive distribution for a left-out observation $n_i$ based on the training set. Coverage is defined by 
\begin{equation*}
\frac{1}{N} \sum 1[n_i \geq l_{c[i]}(t[i])] 1[n_i < r_{c[i]}(t[i])]
\end{equation*}
where $N$ is the the total number of left-out observations considered and $l_{c[i]}(t[i])$ and $r_{c[i]}(t[i])$ the lower and upper bounds of the predictions intervals for the $i$th observation. Coverage at the 80, 90 and 95 per cent levels were considered. 

The validation measures were calculated for 100 sets of left-out observations, where each set consisted of a random sample of one left-out observation per country. Table \ref{lodata} shows the median and standard deviation of each validation measure. The median absolute relative error between the observations and estimated value was less than 10 per cent, and the prediction intervals roughly correspond to a similar coverage of left out observations.

\begin{table}[h!]
\centering
\caption{Validation measures, left-out data}
\label{lodata}
\begin{tabular}{l|r|r}
                        & \multicolumn{1}{l|}{Median} & \multicolumn{1}{l}{Std. Dev} \\ \hline
Absolute relative error & 0.086                       & 0.024                        \\
80\% coverage           & 0.788                       & 0.267                        \\
90\% coverage           & 0.878                       & 0.216                        \\
95\% coverage           & 0.926                       & 0.178                       
\end{tabular}
\end{table}

A similar set of validation measures was calculated comparing the model estimates based on the training data set with the model estimates based on the full data set. Results in Table \ref{modelcomp} are reported for estimates up to (and including) 2005, and post 2005. Model performance is better prior to 2005. This is due to the most recent data being removed, so the data prior to 2005 would be very similar between training and test sets. However the post-2005 measures show estimates are reasonably consistent from using the reduced and full datasets.
\begin{table}[h!]
\centering
\caption{Validation measures, model comparison}
\label{modelcomp}
\begin{tabular}{l|r|r}
                        & \multicolumn{1}{l|}{$\leq 2005$} & \multicolumn{1}{l}{$>2005$} \\ \hline
Absolute relative error &   0.05                               & 0.09                        \\
80\% coverage           & 0.90                             & 0.77                        \\
90\% coverage           & 0.94                             & 0.84                        \\
95\% coverage           & 0.96                             & 0.90                       
\end{tabular}
\end{table}

\section{Discussion}

A new model was introduced for estimating neonatal mortality rates. The model can be expressed as the product of an overall relationship with U5MR and a country-specific effect. The overall relationship with U5MR is a simple linear function, while the country-specific effect is modeled through B-spline regression as a country-specific intercept plus fluctuations around that intercept. Estimates of the NMR were produced for 195 countries, spanning at least the period 1990--2015.

The model appears to perform reasonably well in a wide variety of situations where extent and type of data available varies. In many developed countries, where VR data series are complete and uncertainty around the data is low, NMR estimates follow the data closely. On the other hand, where there is limited data available or if uncertainty around the data is high, estimates are more influenced by the trends in U5MR. 

Model estimates were compared to estimates from the existing IGME model. The notable advantage of this model is that trends in NMR for countries without VR data are driven by the data itself, rather than just reflecting trends in U5MR, as is the case with the existing model. Another advantage of this model is that it is along the same methodological lines as the current model used by IGME to estimate U5MR (Alkema and New, 2014). 

There are several avenues worth investigating further research. The choice of a linear function with changing slope for $f(U_{c,t})$ was a data-driven decision, based on the observed relationship in Figure \ref{scatter}. It would be interesting to compare the performance of models which have a functional form that draws upon existing demographic models. For example, an extended version of the Brass relational logistic model (Brass 1971) and Siler models (Siler 1983) can be used to predict survival in the first months of life, as a function of the survivorship at older ages. 

The potential for bias in estimates from survey data is always a concern. Bias may occur from interviewing a sample that is not representative of the overall population, from selective omission of answers, and can even be influenced by the length of the survey administered (Curtis 1995; Bradley 2015). The data model included an estimation of an overall level of non-sampling error for each survey type, which may account for some biases. However, there is scope to further extend the data model to try to better estimate potential bias in survey data estimates.

The focus of this paper was on the methodology. Future work will also focus on interpretation of results. More investigation is needed on what is potentially causing NMR to be higher- or lower-than expected in outlying countries and whether these are real effects or artifacts of data issues. This distinction is an important one and will become even more so as the focus on child mortality continues to shift towards the early months of life.

\newpage
\section{References}

\bibliographystyle{chicago}
\bibliography{CM}

\begin{description}
\item Alkema, L, and New, JR (2014). `Global estimation of child mortality using a Bayesian B-spline bias-reduction method.' The Annals of Applied Statistics 8(4): 2122--€"2149.
\item Bhutta ZA, Chopra M, Axelson H, Berman P, Boerma T, Bryce J, Bustreo F, Cavagnero E, Cometto G, Daelmans B, de Francisco A, Fogstad H, Gupta N, Laski L, Lawn J, Maliqi B, Mason E, Pitt C, Requejo J, Starrs A, Victora CG, Wardlaw T (2010). `Countdown to 2015 decade report (2000--10): taking stock of maternal, newborn, and child survival.' The Lancet, 375, 2032--2044.
\item Bradley, SEK (2015). `More Questions, More Bias? An Assessment of the Quality of Data Used for Direct Estimation of Infant and Child Mortality in the Demographic and Health Surveys.' Paper presented at PAA 2015. Available at: \texttt{http://paa2015.princeton.edu/uploads/152375}.
\item Brass, W (1971). `On the Scale of Mortality.' In: Brass, W (ed.). Biological Aspects of Demography. London: Taylor \& Francis.
\item Centre for Research on the Epidemiology of Disasters (CRED) (2012). EM-DAT: The CRED International Disaster Database. Available at: \\\texttt{ http://www.emdat.be/database}. 
\item Currie, ID and Durban, M. (2002). `Flexible smoothing with P-splines: A unified approach.' Statistical Modelling, 4: 333--349. 
\item Demographic and Health Surveys (DHS) (2012).  Guide to DHS Statistics: Demographic and Health Surveys Methodology. Available at: \url{http://www.dhsprogram.com/pubs/pdf/DHSG1/Guide_to_DHS_Statistics_29Oct2012_DHSG1.pdf}.
\item Eilers, PHC and Marx, BD (1996). `Flexible smoothing with B-splines and penalties.' Statistical Science, 11 89--121. 
\item Gelman, A and Rubin, D (1992). `Inference from iterative simulation using multiple sequences.' Statistical Science, 7: 457--511.
\item Girosi, F and King, (2007). `Demographic Forecasting.' New Jersey: Princeton University Press. 
\item Lawn J, Cousens S, Zupan J and the Lancet Neonatal Survival Steering Team (2004). `4 million neonatal deaths: When? Where? Why?' The Lancet, 365, 891--900. 
\item Liu L,  Johnson HL, Cousens S, Perin J, Scott S, Lawn J, Rudan I, Campbell H, Cibulskis R, Li M, Mathers C, Black RE (2012). `Global, regional, and national causes of child mortality: an updated systematic analysis for 2010 with time trends since 2000.' The Lancet, 379, 2151--2161. 
\item Lozano R, Wang H, Foreman KJ, Rajaratnam JK, Naghavi M, Marcus JR, Dwyer-Lindgren L, Lofgren KT, Phillips D, Atkinson C, Lopez AD, Murray CJL (2011). `Progress towards Millennium Development Goals 4 and 5 on maternal and child mortality: an updated systematic analysis.' The Lancet, 378, 1139--1169.
\item Mahy, M (2003). `Measuring child mortality in AIDS-affected countries.' Paper presented at the workshop on HIV/AIDS and adult mortality in Developing countries. Available at: \url{http://www.un.org/esa/population/publications/adultmort/UNICEF_Paper15.pdf}.
\item Oestergaard, M.Z., Inoue, M, Yoshida, S., Mahanani, W.R., Gore, F.M., Cousens, S., Lawn, J.E, and Mathers, C.D., (2011). `Neonatal Mortality Levels for 193 Countries in 2009 with Trends since 1990: A Systematic Analysis of Progress, Projections, and Priorities'. PLoS Medicine. DOI: 10.1371/journal.pmed.1001080
\item Pedersen, J and Liu, J (2012). `Child mortality estimation: appropriate time periods for child mortality estimates from full birth histories.' PLoS Medicine. DOI:10.1371/journal.pmed.1001289
\item Plummer, M. (2003). `JAGS: A Program for Analysis of Bayesian Graphical Models Using Gibbs Sampling.' In Proceedings of the 3rd International Workshop on Distributed Statistical Computing (DSC 2003), March 20-22, Vienna, Austria. ISSN 1609-395X. Available at: \url{http://mcmc-jags.sourceforge.net/}.
\item Price M, Klingner J, Ball P (2013) `Preliminary Statistical Analysis of Documentation of Killings in Syria. United Nations Office of the High Commissioner
for Human Rights (OHCHR) technical report. Available at: \texttt{http://www.ohchr.org/Documents/Countries/SY/\\PreliminaryStatAnalysisKillingsInSyria.pdf} 
\item Raftery, AE, Li, N, Sevcikova, H, Gerland, P, and Heilig, GK (2012). `Bayesian probabilistic population projections for all countries.' Proceedings of the National Academy of Sciences of the USA, 109, 13915--13921.
\item Schmertmann, C, Zagheni, E, Goldstein, J and Myrskyla, M (2014). `Bayesian Forecasting of Cohort Fertility.' Journal of the American Statistical Association, 109 (506): 500--513.
\item Siler W (1983). `Parameters of Mortality in Human Populations with Widely Varying Life Spans.' Statistics in Medicine, 2: 373-380.
\item United Nations (UN) (2015). Millennium Development Goal Progress Report. Available at: \texttt{http://www.un.org/millenniumgoals/news.shtml}.
\item United Nations AIDS (UNAIDS) (2014). UNAIDS Gap Report. Available at: \url{http://www.unaids.org/en/resources/campaigns/2014/2014gapreport/gapreport/}
\item United Nations Inter-agency Group for Child Mortality Estimation (IGME) (2015). Levels \& Trends in Child Mortality: Report 2015. Available at: \url{http://childmortality.org/files_v20/download/IGME%20report%202015%20child%20mortality%20final.pdf}
\item United Nations Population Division (UNPD) (2012). World Population Prospects: The 2012 edition. Available at: \texttt{http://esa.un.org/wpp/}.
\item Walker N, Hill K, Zhao, F (2012). `Child Mortality Estimation: Methods Used to Adjust for Bias due to AIDS in Estimating Trends in Under-Five Mortality',  PLoS Medicine 9(8). DOI:10.1371/journal.pmed.1001298.
\item  World Health Organization (WHO). `WHO methods and data sources for global causes of death 2000--2011 (Global Health Estimates Technical Paper WHO/HIS/HSI/GHE/2013.3).' Available at: \url{http://www.who.int/healthinfo/statistics/GHE_TR2013-3_COD_MethodsFinal.pdf}

\end{description}

\appendix

\newpage

\section{Model summary} 
\label{modelsumm}
The full model is summarized below.
\begin{eqnarray*}
r_{c,i} &\sim& N(R_{c,t[c,i]}, \delta_i^2)\\
\delta_i^2 &=& \begin{cases} \tau_{c,i}^2 &\text { for VR data},\\
						\nu_{c,i}^2 + \omega_{s[c,i]}^2 &\text { for non-VR data}\end{cases}\\ 
R_{c,t}&=& f(U_{c,t}) \cdot P_{c,t}\\
\log(f(U_{c,t}))&= &\beta_0 + \beta_1 \cdot \left(\log(U_{c,t}) - \log(\theta)\right)_{[U_{c,t}>\theta]}\\
\log(P_{c,t}) &=& \sum_{k=1}^{K_c} B_{k}(t)\alpha_{c,k}\\
\alpha_{c,k} &=& \lambda_c + [\boldsymbol{D}_{K_c}'(\boldsymbol{D}_{K_c}\boldsymbol{D}_{K_c}')^{-1}\boldsymbol{\varepsilon}_c]_k\\
\lambda_c &\sim& N(0, \sigma_{\lambda }^2)\\
\varepsilon_{c,q} &\sim& N(0, \sigma_{\varepsilon_c}^2)\\
\log(\sigma_{\varepsilon_c}^2) &\sim& N(\chi, \psi^2)\\
\end{eqnarray*}
where

\begin{itemize}
\item $R_{c,t}$ is the true ratio in country $c$ at time $t$,  $R_{c,t}= \frac{N_{c,t} }{U_{c,t} - N_{c,t}}$, where $N_{ct}$ and $U_{c,t}$ are the NMR and U5MR for country $c$ at time $t$, respectively.
\item $r_{c,i}$ is observation $i$ of the ratio in country $c$.
\item $\tau_{c,i}$ is the stochastic standard error, $\nu_{c,i}$ is the sampling error and $\omega_{s[c,i]}^2$ is non-sampling error for series type $s$.
\item $\beta_0$ is global intercept; $\beta_1$ is global slope with respect to U5MR; $\theta$ is the level of U5MR at which $\beta_1$ begins to act.
\item $P_{c,t}$ is country-specific multiplier for country $c$ at time $t$.
\item $B_{k}(t)$ is the $k$th basis spline evaluated at time $t$ and $\alpha_{c,k}$ is splines coefficient $k$.
\item $\lambda_c$ is the splines intercept for country $c$.
\item ${\varepsilon}_{c,q}$ are fluctuations around the country-specific intercept.
\item $\sigma_{\varepsilon_c}^2$ is country-specific smoothing parameter, modeled hierarchically on the log-scale with mean $\chi$ and variance $\psi^2$.
\end{itemize}

The model was fit in a Bayesian framework. Priors are given by
\begin{eqnarray*}
\omega &\sim& U(0,40)\\
\beta_0 &\sim& N(0,100)\\
\beta_1 &\sim& N(0,100)\\
\theta &\sim& U(0,500)\\
\sigma_{\lambda c} &\sim& U(0, 40)\\	
\chi &\sim& N(0, 100)\\
\psi &\sim& U(0,40)
\end{eqnarray*}

\section{Other aspects of the method}

\subsection{Stochastic errors for VR model}
Recall that the observed ratio $r_{c,i}$, which refers to the $i$-th observation of the ratio in country $c$,  is expressed as a combination of the true ratio and some error, i.e.\
\begin{eqnarray}
\label{overallmodel} \nonumber
r_{c,i} &=& R_{c,t[c,i]} \cdot \epsilon_{c,i}\\\nonumber
\implies \log(r_{c,i}) &=& \log(R_{c,t[c,i]}) + \delta_{c,i}
\end{eqnarray}
for $c=1,2, \hdots , C$ and $i=1,\hdots , n_{c}$, where $C=195$ (the total number of countries) and $n_{c}$ is the number of observations for country $c$. The index $t[c,i]$ refers to the observation year for the $i$-th observation in country $c$, $\epsilon_{c,i}$ is the error of observation $i$ and $\delta_{c,i} = \log(\varepsilon_{c,i})$.

For VR data series, the error term $\delta_{c,i}$
 is modeled as 
\begin{equation*}
\delta_{c,i} \sim N(0,  \tau_{c,i}^2 ), 
\end{equation*}
where  $\tau_{c,i}^2$ is the stochastic standard error. These can be obtained once some standard assumptions are made about the distribution of deaths in the first month of life. We assume that deaths in below age five $d_5$ are distributed 
\begin{equation*}
d_5 \sim Pois(B \times _{5}q_0)
\end{equation*}
where $B= $ live births and $_{5}q_0=$ the probability of death between ages 0 and 5. Additionally, we assume deaths in the first month of life $d_n$ are distributed 
\begin{equation*}
d_n \sim  Bin(d_5, p)
\end{equation*}
where $p =$$ _{n}q_0/ _{5}q_0$ and $_{n}q_0$ is the probability of death in the first month of life. Note that the values of $_{n}q_0$ and $_{5}q_0$ come from the raw data. 

The stochastic error was obtained via simulation. For each year corresponding to observation $i$ in country $c$,
\begin{itemize}
\item A total of 3,000 simulations of under-five deaths $d_5$ were drawn from a Poisson distribution $d_5^{(s)} \sim Pois(B \times _{5}q_0)$;
\item A total of 3,000 simulations of neonatal deaths $d_n$ were drawn from a Binomial distribution $d_n^{(s)} \sim Bin(d_5^{(s)}, p)$; 
\item The ratio $y^{(s)} = logit \left( \frac{d_n^{(s)}}{d_5^{(s)}} \right) $ was calculated for each of the simulated samples and the standard error $\tau_{c,i}$ was calculated as $\sigma(\boldsymbol Y)$ where $\boldsymbol Y = (y^{(1)}, y^{(2)},... y^{(s)})$, $s = 3,000$.
\end{itemize}

\subsubsection{SVR data}
For SVR data, the value for the sampling error was imputed based on the sampling error for $U_{c,t[c,i]}$ SVR data, and the observed ratio between the stochastic error of $r_{c,i}$ and the stochastic error of $U_{c,t[c,i]}$. On average, the stochastic error of $r_{c,i}$ was twice as large as the stochastic error of $U_{c,t[c,i]}$. In addition, the sampling error for $U_{c,t[c,i]}$ SVR data was assumed to be 10 per cent. As such a value of 20 per cent was imputed for the sampling error for $r_{c,i}$ SVR data. 

\subsection{Projection} \label{projection}
NMR estimates need to be produced up until 2015, but no data are available up to this year, and many countries also had other recent years missing. As such, country trajectories needed to be projected forward to the year 2015. 

The parameters $\beta_0$, $\beta_1$ and $\theta$, which make up the expected relation with $U_{c,t}$, are fixed over time, as is the country-specific intercept, $\lambda_c$. The component that needs to be projected is the random fluctuations part, the $\varepsilon_{c,q}$ in equation \ref{alphas}. Above, these $\varepsilon_{c,q}$ were assumed to be normally distributed around zero, with some variance $\sigma_{\varepsilon c}^2$. This assumption is used to project the $\varepsilon_{c,1}$ (and thus the splines) forward. 

Start at the first $\alpha_{c,k}$ that is past the last year of observed data. For each time period to be projected: 
\begin{itemize}
\item Draw $\varepsilon_{c,k} \sim N(0, \sigma_{\varepsilon c}^2)$ to obtain $\alpha_{c,k} = \varepsilon_{c,k} + \alpha_{c,k-1} $
\item Repeat to generate $\alpha_k$ for $k$ up to $K_c$, where $K_c$ is knot number needed to cover the period up to 2015.
\end{itemize}
The simulated $\varepsilon_{c,k}$ are generally close to zero, so the method essentially propagates the level of the most recent $\alpha_{c,k}$ that overlaps with the data period with the slope of the expected trajectory, as determined by $f(U_{c,t})$. The projection exercise is necessary in order to maintain a consistent level of uncertainty in the estimates. 

\subsection{Recalculation of VR data for small countries} \label{VRrecalc}
There are several island nations and other small countries that have vital registration data available to calculate NMR. However, observations from these small countries are prone to large stochastic error, which can create erratic trends in NMR over time.

\begin{figure}[h!]
\begin{subfigure}[b]{0.47\textwidth}  \includegraphics[width=\textwidth]{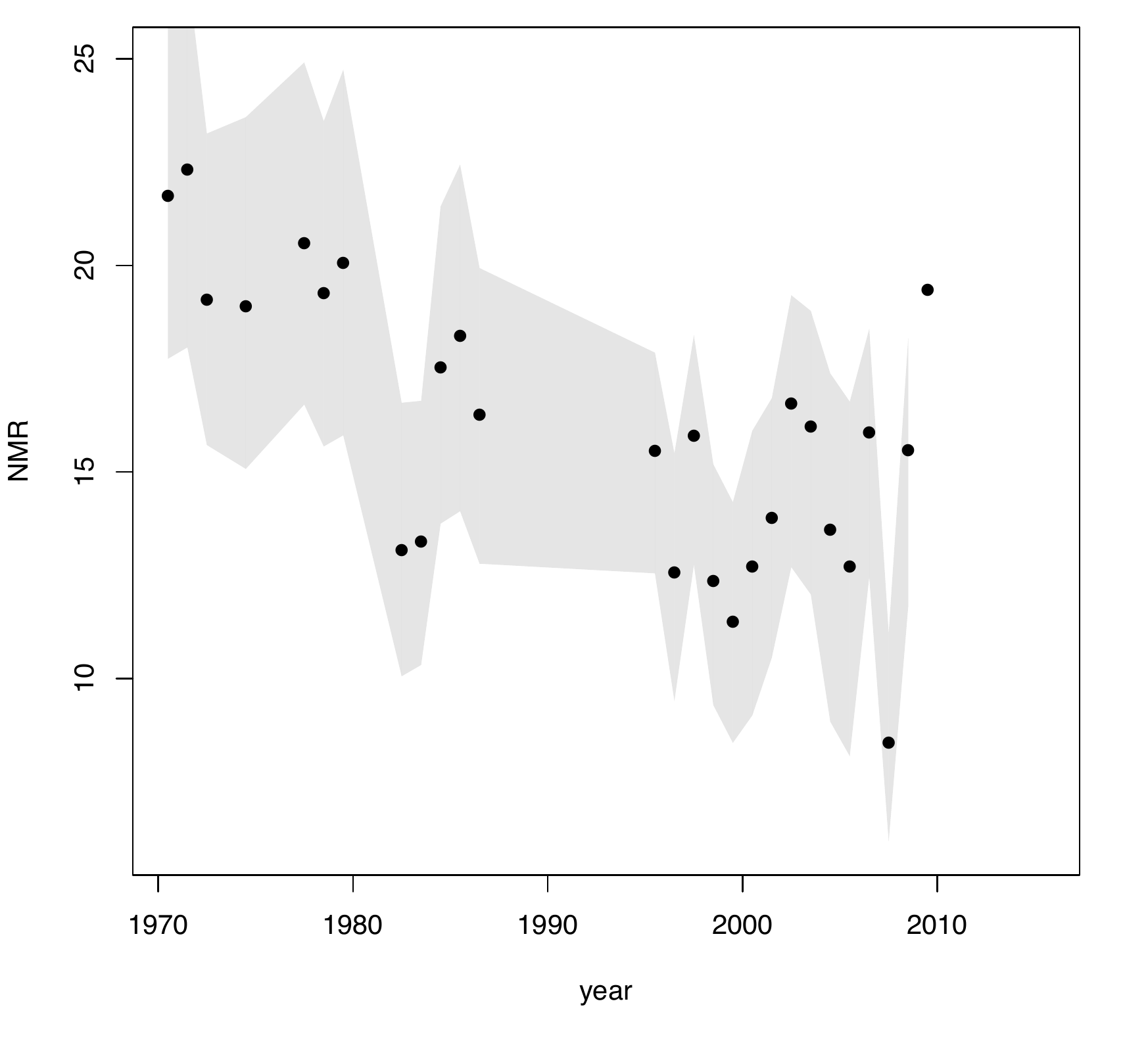} \caption{Original data} \label{fig:VCTorig} 
\end{subfigure} ~ 
\quad 
\begin{subfigure}[b]{0.47\textwidth}  \includegraphics[width=\textwidth]{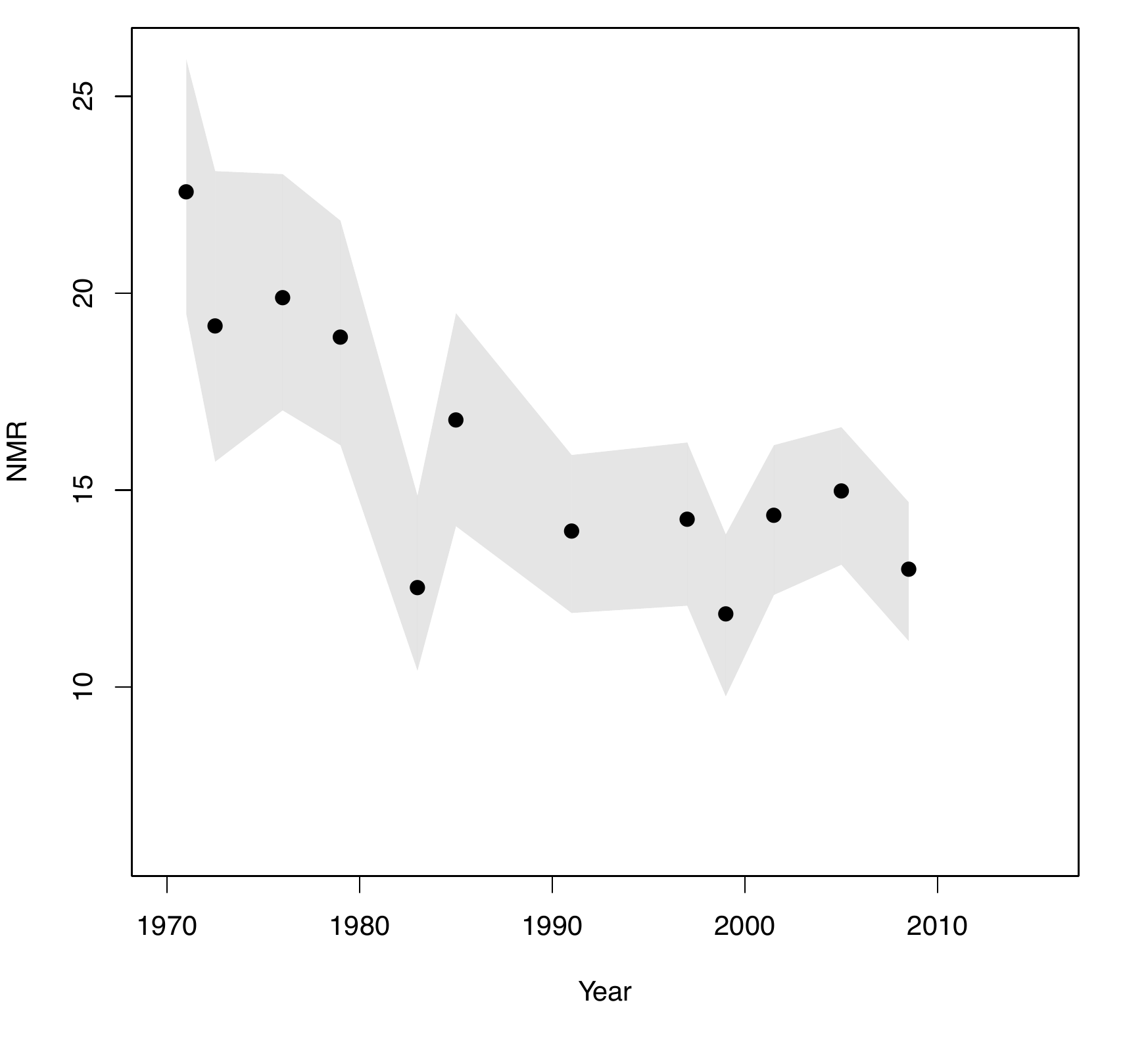} \caption{Recalculated data} \label{fig:VCTrecomb} 
\end{subfigure} ~ 

\caption{Recalculation of VR data: Saint Vincent and the Grenadines}
\label{VCTrecalc}
\end{figure}

To help avoid this issue, observations from adjacent time periods in a particular country are recombined if the coefficient of variation of the observation is greater than 10 per cent. The result is a smaller set of observations with smaller standard errors which display a smoother trend. Figure \ref{VCTrecalc} shows the example of Saint Vincent and the Grenadines on which this process was applied.

NMR is recalculated using the original NMR observations and annual number of live births. For two adjacent years that are to be recalculated:
\begin{itemize}
\item The number neonatal deaths in each year is first calculated as NMR $\times$ live births. 
\item The combined NMR for the two years is then total neonatal deaths divided by total births over the two years. 
\item The standard error of the new NMR estimate is then recalculated based on the process described in \ref{VRdatamodel}.
 \end{itemize}
 After recalculation, the coefficient of variation is calculated for the new estimate. If it is still $>$ 10 per cent, the NMR is recalculated again, recombining with the previous adjacent year. 

\subsection{Crisis deaths}
For some countries, there are known natural or political crises that have caused an excess of deaths; for example the Rwandan genocide or, more recently, the Haiti earthquake and conflict in Syria. For the crisis years, the survey data is unlikely to be representative of the actual number of deaths. 

Adjustments were made to the relevant crisis country-years, using estimates complied by the World Health Organization (WHO). The WHO uses external data sources on the number of deaths, including the Centre for Research on the Epidemiology of Disasters International Disaster Database (CRED 2012) and estimates from UN Office of the High Commissioner for Human Rights for the Syrian conflict (Price 2013). The WHO estimates the proportion of deaths that occur under the age of five (WHO 2013). From there, the best guess of the number of crisis deaths that occur within the first month is simply 1/60th of the total deaths under five years. 

Estimation of crisis countries was firstly done without any crisis adjustments. In addition, the global relation with U5MR, $f(U_{c,t})$, is fit to crisis-free $U_{c,t}$ estimates. The relevant adjustments to country-years were then made post estimation. This was to ensure that the crisis deaths, which are specific to particular years, do no have an effect on the splines estimation.

\subsection{HIV/AIDS countries}
Although there have been vast improvements in recent years, many countries in Sub-Saharan Africa still suffer from relatively high levels of HIV/AIDS-related deaths. This has a substantial effect on the child mortality -- if children living with HIV are not on antiretroviral treatment, a third will not reach their 1st birthday, and half will not reach their 2nd birthday (UNAIDS 2014). However, it is unlikely that children with HIV will die within the neonatal period, and so HIV/AIDS itself does not have and explicit effect on the NMR (although there may be indirect effects on mortality, for example through losing their mother to HIV) (Mahy 2003). 

Due to this disproportionate effect of HIV/AIDS on U5MR compared to NMR, there are several adjustments made to the data used in the model, which leads to NMR being modeled as a function of `HIV-free' U5MR. Firstly, the U5MR data used in the ratio observations are adjusted to incorporate reporting bias. This adjustment accounts for the higher maternal mortality, which leads to under-estimation of child mortality from surveys (Walker et al.\ 2012). Once adjusted, the HIV/AIDS deaths are removed from U5MR, using estimates of deaths provided by UNAIDS (UNAIDS 2014). The result is a ratio of neonatal to other child mortality which is free of HIV deaths. In addition, the global relation with U5MR, $f(U_{c,t})$, is fit to HIV-free data. Unlike the crisis adjustments, no HIV deaths were added in post-estimation, because it is assumed no neonatal deaths are due to HIV/AIDS.

\subsection{Countries with no data}
There were twelve UN-member countries for which the IGME produces NMR estimates for, but where there are no available data. For these countries, the estimates of NMR are based on the global relation with U5MR, $f(U_{c,t})$. Additionally, some steps are needed to obtain the appropriate uncertainty around these estimates. For country $c$:
\begin{itemize}
\item Draw $\lambda_c \sim N(0, \sigma_{\lambda}$);
\item Set $\alpha_1 = \lambda_c$;
\item Draw $\varepsilon_1 \sim N(0, \sigma_{\varepsilon})$; where $\sigma_{\varepsilon} = e^{\chi}$ is the global smoothing parameter, based on equation \ref{smoothingequation};
\item Set $\alpha_2 = \alpha_1 + \varepsilon_1$;
\item Repeat to generate $\alpha_k$ for $k=3, \dots , K_c$. $K_c$ is the number of spline knots needed to cover the period 1990--2015.
\end{itemize}

\end{document}